\documentclass[a4paper,table,11pt]{article}
\usepackage{jheppub} 

\usepackage{amsmath}
\usepackage{amsfonts}
\usepackage{bm}
\usepackage{mathrsfs}
\usepackage{dsfont}
\usepackage{subfig} 
\usepackage{empheq}
\usepackage{esint}
\usepackage{hyperref}
\usepackage[shortlabels]{enumitem}
\usepackage{tikz, pgf} 
\usetikzlibrary{arrows}
\usepackage{xcolor}
\usepackage{pgfplots}
\usepackage{natbib} 
\usepackage{braket}
\usepackage{slashed}
\usepackage{bbold}

\usepackage[utf8]{inputenc}
\usepackage{array}
\usepackage{ctable}

\newcommand{\ob}[1]{\ensuremath{\overline{\raisebox{0pt}[1.10\height]{#1}}}}
\newcommand{\overbar}[1]{\ob{${#1}$}}

\newcommand{\fr}[2]{\frac{#1}{#2}}
\newcommand{\tc}{t_{\mathcal{C}}}

\newcommand{\del}{\partial}
\newcommand{\dl}{\delta}
\newcommand{\w}{\omega}

\newcommand{\be}{\beta}

\newcommand{\calNm}{\mathcal{N}_{m}}

\newcommand{\Ctor}{\zeta}

\newcommand{\dCtor}{\fr{d}{d\Ctor}}

\newcommand{\FD}{n_{\omega}^{\text{FD}}}

\newcommand{\Go}[1]{\:\Gamma^{(#1)}}

\newcommand{\ProCtor}{\mathcal{P}^{\Ctor}}

\newcommand{\psiPb}{\psi_{\bar{P}}}
\newcommand{\psiFb}{\psi_{\bar{F}}}

\begin{document}

\title{Fermionic Open EFT from Holography}
\author[a]{R. Loganayagam,}
\author[b]{Krishnendu Ray,}
\author[a]{Akhil Sivakumar.}

\emailAdd{nayagam@icts.res.in}
\emailAdd{krishnendu.ray@physics.ox.ac.uk}
\emailAdd{akhil.sivakumar@icts.res.in}
\affiliation[a]{ International Centre for Theoretical Sciences (ICTS-TIFR),
Tata Institute of Fundamental Research,
Shivakote, Hesaraghatta,
Bangalore 560089, India.}
\affiliation[b]{Rudolf Peierls Centre for Theoretical Physics, University of Oxford, \\ Parks Road, Oxford, OX1 3PU, United Kingdom.}
\begin{abstract}
{In this work we initiate the study of open effective field theories of fermions interacting with  holographic baths. As a first step in this direction, we explain how the recently identified holographic Schwinger--Keldysh saddles naturally reproduce the Fermi--Dirac statistics of boundary correlators. We study Dirac fermions propagating in the background of a doubled AdS$_{d+1}$ Schwarzchild blackbrane. We solve for both the ingoing (quasi--normal) as well as outgoing (Hawking) modes of the Dirac field in a boundary gradient expansion and compare our findings against exact results in $d=2$.  The holographic Schwinger--Keldysh boundary conditions automatically select a unique combination of these modes consistent with fermionic KMS relations. Evaluating the bulk on--shell action using these solutions, we obtain an explicit expression for the influence phase of the probe fermion in a gradient expansion.}
\end{abstract}
\maketitle

\section{Introduction}\label{Introduction}

Several strongly correlated electronic systems have been reported to show novel metallic behaviour \cite{Haldane_1981,Gegenwart_2008,Sachdev_2016}. Their low energy physics cannot be captured by quasiparticle excitations. These systems thus fall outside the standard Fermi--liquid paradigm  and are hence termed non--Fermi liquids. A distinct feature of these novel metals is that many of them seem to undergo critical phase transitions at low temperatures, and are thus governed by the fermionic cousins of the bosonic 
quantum phase transitions \cite{sachdev_2011}. This motivates a plausible origin of their unusual properties --- the fermionic excitations in these systems are thought to be interacting with a conformal sector leading to a  quick decay of quasiparticles and a breakdown of Fermi--liquid structure \cite{Varma_2002,Mross_2010,Lee_2009}.
While many examples of this type are known and well--studied, an overarching theoretical framework is not yet available.

Many fundamental questions about non--Fermi liquids remain unanswered. What is the most efficient description of their long distance dynamics? What is the best characterisation of their low energy transport \cite{Stewart:2001zz,RevModPhys.78.743}?  We propose that a useful  point of view to characterise strongly correlated metallic phases of matter is  via the effective theory of probe fermions propagating in these systems. After all, a universal experimental feature uniting materials with or without quasiparticles is the fact that they can be probed by injecting electrons into them (say in STM/ARPES type measurements). Since such probe electrons often suffer inelastic scattering, decoherence and dissipation as they move through the material,  the effective theories for such probes  should be understood from the perspective of open effective field theories (open EFTs). Open EFTs describe the effective dynamics of a collection of fields of interest (the system) interacting with a  set of inaccessible  fields (the bath), after the latter fields are integrated out. Developing such a systematic framework is especially necessary if we have to go beyond the lower point correlations (which characterise conductivity etc.) and to include the physics of out--of--time ordered correlations.

In most cases, deriving open effective actions from a unitary microscopic theory is a tricky exercise. This is because one usually lacks a good way to separate out the bath and system degrees of freedom. To have a local effective action at some time scale \(\tau_{\text{eff}}\), it is necessary that the correlators of the bath relax at a time scale  \(\tau_{\text{bath}} < \tau_{\text{eff}}\). This accords with the intuition that the bath should `forget' fast enough to let the system evolve in a Markovian or local fashion. For generic baths, \(\tau_{\text{bath}} \) decreases as the interaction coupling increases; stronger coupling implies a faster mixing of information and thus a faster decay of correlations. This indicates that non--Fermi liquids can be thought of as Markovian baths --- they lack long--lived quasiparticles and have a characteristic relaxation time of the order of the inverse temperature, $\beta$ \cite{sachdev_2011,article}. Therefore at time scales larger than $\beta$, the probe fields interacting with them are described by local effective actions. When long--lived excitations are present they show up
as memory effects in the open EFT of the probe. 

 As mentioned above, the fermionic fields in novel metals are thought to be interacting with some critical sector of the metallic system, leading to the the break down of the conventional Fermi liquid description. A simple way to model such interactions is to couple free fermions to a critical theory, like gapless bosons \cite{PhysRevB.14.1165,PhysRevB.48.7183,RevModPhys.79.1015}.  A major difficulty with such models is that one has to carefully devise an expansion parameter giving a controlled approximation of the strongly coupled dynamics, without losing its essential features giving rise to the non--Fermi liquid behaviour. 

Another class of toy models widely used to study strongly coupled critical systems is based on holographic CFTs. Holographic models offer a great advantage over the field theoretical ones as the former come with a natural perturbative description in terms of the dual gravitational dynamics. The finite temperature states of these models are dual to black brane backgrounds which dissipate and relax at a time scale of order $\beta$ \cite{Horowitz_2000}. This in turn fixes the relaxation time scale of the field theory, making them qualitatively similar to the novel metals discussed above. The application of holographic techniques to fermionic systems  have uncovered a rich variety of phases of  Fermi liquid and non--Fermi liquid kinds  \cite{hartnoll2018holographic,zaanen_liu_sun_schalm_2015}. Recent developments in real time holography (which we review below) make these holographic models
a good avenue to develop the open EFT perspective described above.

With these motivations, we  initiate a study of fermionic  open EFTs using holography. The setup we consider is similar to the semi--holographic model pioneered by \cite{Faulkner_2011}: the microscopic action consists of an external probe fermionic field $\psi$ which couples to a holographic CFT at a finite temperature via  a fermionic, single trace primary $\mathcal{O}$.  The open effective action for $\psi$ fields is then derived by integrating out the CFT bath in the real time Schwinger--Keldysh path integral \cite{Feynman:1963fq,Schwinger:1960qe} (see \cite{Rammer:2007zz,Bellac:2011kqa,kamenev_2011} for a textbook description). The effect of bath fields are then encoded in the \emph{influence phase} \cite{Feynman:1963fq} added to the  Schwinger--Keldysh effective action of $\psi$. Our goal is to compute
this influence phase holographically. The fermionic single trace primary $\mathcal{O}$ is dual to a bulk Dirac field $\Psi$, and the 
influence phase comes from evaluating the Dirac action on a field configuration of $\Psi$ probing a `gravitational Schwinger--Keldysh'
(grSK) geometry \cite{Skenderis:2008dg,Skenderis:2008dh,vanRees:2009rw,Glorioso:2018mmw,Chakrabarty:2019aeu,Jana:2020vyx}.

We begin by reminding the reader that we still lack a systematic formalism to compute real time SK path integrals in AdS/CFT.
Over the last two decades, many prescriptions have been proposed, gradually expanding the class of real time observables that
can be computed holographically. The pioneering work of Son and Starinets \cite{Son:2002sd} reduced the computation of the
retarded/causal Green's functions in an equilibrium CFT to ingoing/quasi--normal field configurations on an AdS black brane. This prescription has now been successfully generalised to near--equilibrium hydrodynamic configurations in the fluid--gravity correspondence 
(see \cite{Rangamani:2009xk, Hubeny:2011hd} and references therein). It has however proved tricky to go beyond the causal
Green's functions to compute other real time observables (encoding the physics of fluctuations). On the gravitational side,
the physical challenge is to account for Hawking radiation and its interactions with the infalling modes. One is forced 
to deal with this physics by analytically continuing from Euclidean or eternal black brane computations \cite{Herzog:2002pc,Son:2009vu}.
A particularly relevant work in the present context is \cite{Giecold:2009tt} where Giecold studied a Dirac field probing
an eternal black brane background. 

While these analytic continuations tricks have by now been successfully extended to compute a small class of out--of--time ordered correlation functions in equilibrium \cite{Shenker:2013pqa}, the vast class of non--equilibrium, real time correlations  computed by generalised SK path integrals are not yet within reach. For this reason, over the past decade many authors have attempted
to formulate a prescription that directly computes real time correlations in holography without obscuring the relevant physics
under analytic continuation gymnastics. The pioneering works in this regard are by Skenderis and van Rees \cite{Skenderis:2008dg,Skenderis:2008dh,vanRees:2009rw} who suggested a doubled geometry that fills out the SK  contour in the CFT (See Fig.\ref{fig:fixedv}). In the case of thermal CFT states, the doubled geometry consists of two Schwarzchild black branes glued together along a common future horizon. See \cite{Leigh:2009eb,Barnes:2010jp,Barnes:2010ev,Botta-Cantcheff:2017qir} for further work in this direction.

The Skenderis--van Rees prescription for the horizon was improved upon by Glorioso, Crossley and Liu in \cite{Glorioso:2018mmw}, giving further clarity to the horizon structure of these geometries. In the updated prescription, the horizon region is replaced by a `horizon cap' interpolating between the two black brane exterior spacetimes. We discuss this geometry more elaborately in $\S$\ref{Section: grSK and open EFTs}. Following \cite{Jana:2020vyx}, we refer to this geometry as \emph{gravitational Schwinger--Keldysh saddle} (grSK).
This Schwinger--Keldysh saddle was used in \cite{Chakrabarty:2019aeu,Jana:2020vyx} to derive interacting open effective actions for Brownian particles and scalar fields probing the CFT,  and was shown to pass a variety of non--trivial checks at the interacting level.  Our goal in this work is to extend these ideas to fermionic fields to construct fermionic open EFTs. 

We will end this introduction by quoting the central physical result of this work followed by an outline. A crucial check that our 
prescription does indeed reproduce the correct real time dynamics of fermionic radiation is the derivation of the Fermi--Dirac
function from the bulk. More precisely, our answers should be consistent with the  fermionic Kubo--Martin--Schwinger \cite{Kubo:1957mj,Martin:1959jp} conditions for SK correlators. Note that this is not intuitive: after all in the bulk we work completely classically and it is non--trivial to derive Bose--Einstein or Fermi--Dirac statistics without any quantisation whatsoever. As explained in
\cite{Chakrabarty:2019aeu,Jana:2020vyx} this works in the bosonic examples as follows: the structure of grSK saddles is such that the  outgoing Hawking radiated modes behave as evanescent waves near the horizon cap undergoing a Boltzmann--like suppression \emph{classically} when compared across the two exteriors. The horizon cap region is however completely transparent to the
ingoing quasi-normal modes. An interference between these two then results in the appearence of the Bose--Einstein function
in the classical calculation. 

As we will argue later in this work, for a field of spin $s$, the classical transmission amplitude for the outgoing modes across the horizon cap is  given by, 
\begin{equation}
e^{-\beta \omega + 2 \pi i s} \, .
\end{equation}
For integer $s$ this reduces to the simple Boltzmann weight, whereas for half integer $s$ we get an additional phase factor of $-1$. This then leads to a destructive interference of the ingoing and the outgoing modes, perfectly reproducing the Pauli exclusion
felt by half--integer fields. The classical physics of grSK saddles hence automatically reproduces the spin--statistics relation
implementing the Fermi--Dirac statistics of the boundary correlators. We see this as a non--trivial check of this prescription.

Here is an outline of the rest of this work: we begin in $\S$\ref{Fermions in SK} with a review of the Schwinger--Keldysh
formalism as applied to fermions, followed by a review of necessary ideas from real time holography  in \S\ref{Section: grSK and open EFTs}. The main results of this work are then summarised in two sections: in  \S\ref{sec:FermionicProp}, we explain in detail the classical origin of Fermi--Dirac statistics when solving Dirac equations in the grSK saddle. In \S\ref{sec:GradientExp}, we give, for the first time, an explicit solution of the Dirac equation in the black--brane background in a derivative expansion. We also describe the structure 
of the ingoing quasi--normal modes and  the outgoing Hawking modes as well as the superposition of the two that satisfies the 
appropriate boundary conditions on the grSK saddle. We conclude with a discussion of future directions in \S\ref{sec:DiscConc}.
Various technical computations are relegated to the appendices including the details about spinors and gamma matrices in 
Appendix \ref{Spinor and Gamma Conventions}. A detailed comparison with the exact solution of the Dirac equation in $d=2$ is presented in Appendix \ref{app:Exact Solution}. Details about the asymptotic expansions of the functions appearing in the derivative expansion
are discussed in Appendix \ref{app:Asymp}, and we end with a table of notation made  for the convenience of the reader.

\section{Fermions in the Schwinger--Keldysh formalism}  \label{Fermions in SK}

In this section, we will review the Schwinger--Keldysh formalism below with an emphasis on fermionic fields (see \cite{Rammer:2007zz,Bellac:2011kqa,kamenev_2011} for a textbook description). 

Consider a unitary field theory which contains fermionic operators collectively denoted as $ \mathcal{O} $. The SK generating functional is given by 
\begin{equation} \label{SK Partition Function}
\begin{split}
\mathcal{Z}_{\mathrm{SK}}\left[\psi,\overbar{\psi} \right]\equiv \textbf{Tr} \left(  U[\psi_R,\overbar{\psi}_R]\ \rho_{\text{init}} \  U[\psi_L,\overbar{\psi}_L] \right)\, ,
\end{split}
\end{equation}
where $\rho_{\text{init}}$ is the initial density matrix and $\psi_{R,L}$ represent the sources for the fermionic operator $\mathcal{O}$  on the right/left contour. Equivalently, we would like to think of $\psi_{R,L}$  as probe fermion fields coupling to the field theory under question. This provides an equivalent interpretation of the above generating function as the \emph{influence functional} for this probe field which encodes the effect of the environment.

The operator $U$ is the unitary evolution operator deformed by the sources $\psi _{R,L}$. When $\rho_{\text{init}}$ is thermal, the above generating function can be regarded as the path integral of the field theory living on a complex time contour \(\mathcal{C}\), as shown  in Fig.\ref{fig:fixedRho}.  The part of the contour indicating forward (backward) evolution in time is labeled \(\mathcal{C}_R\) (\(\mathcal{C}_L\)). \(\mathcal{C}_E\) corresponds to an Euclidean evolution of length \(\be \) and is a surrogate for the thermal circle. This part of the contour prepares the thermal initial state. 

When a Lagrangian description is available, we can write down a  SK action defined by,
\begin{equation} \label{SKAction}
\begin{split}
 \mathds{S}_{\text{SK}} \equiv \ointctrclockwise\displaylimits_{\mathcal{C}} \; \mathrm{d}\tc \;  \mathrm{d}^{d-1}\bm{x} \ \mathcal{L}[\mathcal{O}] \, , \\
\end{split}
\end{equation}
where \(\tc\) stands for the contour time on  \(\mathcal{C}\) and  \( \mathcal{L}[\mathcal{O}] \) is  the Lagrangian density of the field theory. The corresponding SK generating functional can be schematically written as,
\begin{equation} \label{SK Partition FunctionLag}
\mathcal{Z}_{\mathrm{SK}}\left[\psi,\overbar{\psi} \right] = \int \mathcal{D}[\mathcal{O},\overbar{\mathcal{O}}] \ \text{exp} \left\{i \:\mathds{S}_{\mathrm{SK}} + i \int\displaylimits_{\mathcal{C}} \mathrm{d}\tc \: \mathrm{d}^{d-1}\bm{x} \; \left(\overbar{\psi} \, \mathcal{O} + \overbar{\mathcal{O}} \, \psi \right) \right\} \, .
\end{equation}
Here $\mathcal{D}[\mathcal{O},\overbar{\mathcal{O}}]$ schematically represents the path integral over the elementary fields out of which the operators $\mathcal{O},\overbar{\mathcal{O}}$ are constructed. 

It is convenient to distinguish between the operators supported on the segments \(\mathcal{C}_{R} \)  and \(\mathcal{C}_{L} \) of the contour. We define,
\begin{equation}
\begin{split}
&\mathcal{O}(\tc,\bm{x})\equiv \  \mathcal{O}_R(t,\bm{x}) \, , \qquad \psi(\tc,\bm{x}) \equiv \psi_R(t,\bm{x}) \, , \qquad  t=\tc \  ; \quad \tc \in \mathcal{M}_R \\
&\mathcal{O}(\tc,\bm{x})\equiv\  \mathcal{O}_L(t,\bm{x}) \, , \qquad \psi(\tc,\bm{x}) \equiv \psi_L(t,\bm{x}) \, , \, \qquad t = 2t_f- \tc \ ; \quad \tc \in \mathcal{M}_L \\ 
&\mathcal{O}_R(t_f,\bm{x}) \equiv \  \mathcal{O}_L(t_f,\bm{x}) \, , \quad \psi_R(t_f,\bm{x}) \equiv \psi_L(t_f,\bm{x}) \, , 
\end{split}
\end{equation}
where $t_f$ is the future most or turning point on the contour. Once we adopt these new variables, the time variable \( t\) can be identified with the physical time \(x^{0}\), and fields  \(\mathcal{O}_R \) and \(\mathcal{O}_L \) are field degrees of freedom corresponding to forward  and backward temporal evolutions respectively; they are identified at \(t_f\). The initial time configurations of these fields are weighted by an Euclidean path integral on the segment \(\mathcal{C}_E \) which corresponds to the preparation of the thermal initial state. This segment of the contour will play a largely passive role in our story as we do not intend to source the fields there\footnote{We will consider only the thermal initial state. Field insertions in the Euclidean segment correspond to driving the thermal state out of equilibrium.}. 

The SK two--point functions are defined by,
\begin{equation}
\begin{split}
  \mathcal{S}^{IJ}(x,y) \equiv \fr{1}{i} \fr{\dl}{\dl \overbar{\psi}_I(x)} i \fr{\dl}{\dl \psi_J(y)} \ln \mathcal{Z}\bigg|_{\psi=\overbar{\psi}=0} \qquad = \ \left(\begin{array}{c c} \mathcal{S}^{RR} & \mathcal{S}^{RL} \\
  \mathcal{S}^{LR} & \mathcal{S}^{LL}
  \end{array}\right) \, ,
\end{split}
\end{equation}
where we follow the convention that Latin letters \(I, J = R, L\)  denote contour indices  and  Greek letters \(\alpha, \be\) are fermionic indices\footnote{We adopt these conventions for this section only. Further, we do not distinguishing between upper and lower indices.}. Since path integrals generate contour ordered correlators, we have, 
\begin{equation}
\begin{split}
  \mathcal{S}^{RR}_{\alpha \beta}(x,y) =&\   \langle \mathds{T}\:\mathcal{O}_{\alpha}(x) \overbar{\mathcal{O}}_{\beta}(y)  \rangle \, , \qquad \mathcal{S}^{RL}_{\alpha \beta}(x,y) = - \:  \langle \overbar{\mathcal{O}}_{\beta}(y) \mathcal{O}_{\alpha}(x)  \rangle \, , \\ 
  \mathcal{S}^{LR}_{\alpha \beta}(x,y) =&\   \langle \:\mathcal{O}_{\alpha}(x) \overbar{\mathcal{O}}_{\beta}(y)  \rangle \, , \qquad \quad  \mathcal{S}^{LL}_{\alpha \beta}(x,y) =  \: \langle \hat{\mathds{T}} \:\mathcal{O}_{\alpha}(x) \overbar{\mathcal{O}}_{\beta}(y) \rangle \, . 
\end{split}
\end{equation}
Here \(\mathds{T}\) and \(\hat{\mathds{T}}\) denote the time ordering and anti--time ordering of the operators. The additional negative sign on \(\mathcal{S}^{RL} \) comes from the Grassmann odd property of fermionic operators.

The SK Green's functions given above simplify further in the so called `Keldysh--rotated' field basis, given by the average and difference of the $R$ and $L$ field variables\footnote{Our convention here differs from that of \cite{kamenev_2011} where the definitions of \(\overbar{\mathcal{O}}_a\) and \(\overbar{\mathcal{O}}_d\) are interchanged.}. 
\begin{equation}
\begin{split}\label{ad Basis}
 \mathcal{O}_a =&\   \fr{1}{2}\left(\mathcal{O}_R+\mathcal{O}_L \right) \, , \qquad \mathcal{O}_d = \mathcal{O}_R- \mathcal{O}_L \, .  \\ 
 \overbar{\mathcal{O}}_a =&\   \fr{1}{2}\left(\overbar{\mathcal{O}}_R+\overbar{\mathcal{O}}_L \right) \, , \quad \ \; \overbar{\mathcal{O}}_d = \overbar{\mathcal{O}}_R- \overbar{\mathcal{O}}_L \, .
\end{split}
\end{equation}
Similarly, we define the average/difference sources,
\begin{equation}
\begin{split}\label{ad Sources}
\psi_a =&\   \fr{1}{2}\left(\psi_R+\psi_L \right) \, , \qquad \psi_d = \psi_R- \psi_L \, . \\ 
 \overbar{\psi}_a =&\   \fr{1}{2}\left(\overbar{\psi}_R+\overbar{\psi}_L \right) \, , \quad \ \; \overbar{\psi}_d = \overbar{\psi}_R- \overbar{\psi}_L \, .
\end{split}
\end{equation}

The Green's function matrix, when expressed in the Keldysh basis becomes,
\begin{equation}
\begin{split}
\mathcal{S}^{aa}_{\alpha \beta}(x,y) =& \ \fr{1}{2}\left\langle\left[\mathcal{O}_{\alpha}(x),\overbar{\mathcal{O}}_{\beta}(y) \right]\right\rangle \, , \qquad  \mathcal{S}^{ad}_{\alpha \beta}(x,y) = \ \theta(x^{0}>y^{0})\left\langle\left\{\mathcal{O}_{\alpha}(x),\overbar{\mathcal{O}}_{\beta}(y) \right\}\right\rangle \, , \\
\mathcal{S}^{dd}_{\alpha \beta}(x,y) =& \ 0 \, , \qquad \qquad \qquad \qquad \qquad \ \, \mathcal{S}^{da}_{\alpha \beta}(x,y) = \ -\theta(x^{0}<y^{0})\left\langle\left\{\mathcal{O}_{\alpha}(x),\overbar{\mathcal{O}}_{\beta}(y) \right\} \right\rangle \, .
\end{split}
\end{equation}
where,
\begin{equation}
\mathcal{S}^{ij}(x,y) \equiv \langle \mathcal{O}_i(x)\overbar{\mathcal{O}}_j(y)\rangle \, , \qquad i,j=a,d \, . 
\end{equation}
From the above we recognise the retarded and advanced Green's functions as,
\begin{equation}
\mathcal{S}^{\mathrm{ret}}_{\alpha \beta}(x,y) \equiv \mathcal{S}^{ad}_{\alpha \beta}(x,y) \, , \qquad \mathcal{S}^{\mathrm{adv}}_{\alpha \beta}(x,y) \equiv \mathcal{S}^{da}_{\alpha \beta}(x,y) \, .
\end{equation}

An important advantage of the Keldysh basis is the statement that the two point function 
$\mathcal{S}^{dd}_{\alpha \beta}(x,y) =0 $.  This can be traced to the fact that the correlations of 
difference operators $\mathcal{O}_{d}$ are generated by setting $\psi_R=\psi_L$ in the original SK generating function \eqref{SK Partition Function}. The underlying unitarity and the cyclicity of trace then imply that the SK generating function reduces to unity, demonstrating that any $n$--point correlation function of only difference operators identically vanishes. This is the \emph{Schwinger--Keldysh collapse rule} obeyed by SK correlators. Since,
\begin{equation}
\overbar{\psi}_R\mathcal{O}_R-\overbar{\psi}_L\mathcal{O}_L =
\overbar{\psi}_a\mathcal{O}_d+\overbar{\psi}_d\mathcal{O}_a\, ,
\end{equation}
this implies that the  generating function of correlations never has a term with only $\psi_a$'s.

The third and last basis set we introduce is the \emph{retarded--advanced (RA)} basis,
\begin{equation} 
\begin{split}
\psi _{\bar{F}} \equiv n^{\mathrm{FD}}_{\omega } \left(\psi _{R} - \psi _{L}\right) - \psi _{R} \, , \qquad \psi _{\bar{P}} \equiv n^{\mathrm{FD}}_{\omega } \left(\psi _{R} - \psi _{L}\right),
\end{split}
\end{equation}
where we have defined the Fermi--Dirac factor, 
\begin{equation} \label{FD factor}
\begin{split}
n_{\omega }^{\mathrm{FD}} = \frac{1}{1 + e^{\beta \omega }} \, .
\end{split}
\end{equation}

The combination above (which differs slightly from the RA basis defined in say \cite{Chou:1984es}) is motivated by the analysis presented in \cite{Chaudhuri:2018ymp}. This 
combination is chosen to make manifest the Kubo--Martin--Schwinger \cite{Kubo:1957mj,Martin:1959jp} relations arising from thermality and the cyclicity of the trace.
We will refer the reader to \cite{Chaudhuri:2018ymp} for a complete 
description of such constraints on real time-correlators. For our purposes,
it is sufficient to simply state the final result: the SK collapse rules and
the KMS conditions together imply that \emph{the  generating function of SK correlations
can never have  term with only $\psi _{\bar{P}}$s or terms with only $\psi _{\bar{F}}$s, i.e., all non-zero terms have at least one $\psi _{\bar{P}}$ and at least one $\psi _{\bar{F}}$} .
As we will see, these conditions arise naturally in our analysis in the bulk. For two point functions in Keldysh basis, these two conditions imply $\mathcal{S}^{dd}=0$, and in addition, \begin{equation}
\begin{split}
\mathcal{S}^{aa} = \left(\fr{1}{2}-\FD  \right) \left( \mathcal{S}^{ad}- \mathcal{S}^{da}\right)\, .
\end{split}
\end{equation}

\section{The grSK geometry and open EFTs} \label{Section: grSK and open EFTs}
As mentioned in $\S$\ref{Introduction}, our goal is to construct an open effective field theory for a fermionic probe field coupled to a thermal environment with a gravitational dual --- we present the framework for this computation in this section.

 We first focus on the holographic thermal bath/environment, detailing a holographic recipe for calculating its real--time correlators. This is done in three steps: in $\S$\ref{Subsection: Fermions in AdS/CFT}, we quickly review the standard prescription for computing response functions for spinors from the AdS/CFT correspondence. We then introduce a complex two--sheeted geometry that is the relevant gravitational dual of the asymptotic Schwinger--Keldysh contour in $\S$\ref{Subsection: grSK geometry}. In $\S$\ref{Subsection: Open EFT from Holography}, we describe how the usual AdS/CFT dictionary can be adapted to this new geometry, allowing us to compute the SK generating function (which as we explained before is equivalent to computing the open effective action   for the fermionic probe).

\subsection{Fermions in AdS/CFT} \label{Subsection: Fermions in AdS/CFT}
The discussion of the AdS/CFT correspondence for spinors that we present here closely follows the original references on the subject  \cite{Henningson:1998cd,Henneaux:1998ch,Mueck:1998iz,Iqbal:2009fd}. A large part of this short review will be carried out in Euclidean space. At the very end, we will describe how to move all our expressions back to the real--time domain, which is the starting point for the discussion in $\S$\ref{Subsection: Open EFT from Holography}.

We begin with the metric of an asymptotically Euclidean AdS$_{d+1}$ spacetime, $\mathcal{M}_{E}$, written out in the (Euclidean) Eddington--Finkelstein coordinates.  We are interested in studying a single free fermion living in this spacetime.
\begin{samepage}
	\begin{equation} 
	\begin{split}
	&\mathrm{d}s^{2} = g_{AB} \ \mathrm{d}x^{A} \, \mathrm{d}x^{B} = g_{v_{E} v_{E}} \, \mathrm{d} v_{E}^{2} + 2 \, i \,  \mathrm{d}v_{E} \, \mathrm{d}r + g_{ij} \, \mathrm{d}{x^{i}} \, \mathrm{d}x^{j} 
	\end{split}
	\end{equation}
	\begin{equation*} 
	\begin{split}
	g_{v_{E} v_{E}} \, \xrightarrow[r \to \infty]{} \ r^{2} \, , \qquad g_{ij} \  \xrightarrow[r \to \infty]{} \ r^{2} \, \delta _{ij}
	\end{split}
	\end{equation*}
\end{samepage}
Here $v_{E}$ is the advanced Euclidean Eddington--Finkelstein time and the AdS radius has been set to unity. In addition to the coordinates and metric, we need to align the Gamma  matrices and spinors that we use to write down the Dirac action in this geometry. This is done using a tetrad (tangent) frame, i.e., a set of $(d+1)$ linearly independent, smooth 1--form fields\footnote{A word on notation: we use the uppercase alphabet $(A, B, \ldots)$ to denote abstract spacetime indices, and the lowercase alphabet $(a, b, \ldots)$ to denote abstract tetrad frame indices. Field theory indices will be denoted by $(\mu , \nu , \ldots)$. Whenever a particular value of a tetrad frame index is referred to, we will enclose it within a parenthesis to distinguish it from the corresponding spacetime index.} $E^{a}_{A}$ that satisfy,
\begin{equation}\label{Tetrad Defintion}
\begin{split}
\delta _{ab} \, E^{a}_{A} \, E^{b}_{B} = g_{AB},
\end{split}
\end{equation}
where $\delta _{ab}$ is the $(d+1)$ dimensional Euclidean metric. The Gamma  matrices and spinor fields are now defined with respect to this tangent frame. Collectively, these tetrads fix the orientations of these objects in the ambient spacetime --- as an important example, the spacetime Gamma  matrices are defined as $\Gamma ^{A} = \Gamma ^{a} \, E_{a}^{A}$. 

From \eqref{Tetrad Defintion} is clear that the choice of tangent frame is hardly unique, with different frames leading to different equations of motion. For now, we leave our choice of tetrads unspecified as our current discussion is independent of its details. 

With these structures defined, the Euclidean action for a minimally coupled Dirac fermion $\Psi $ propagating in this spacetime is,
\begin{equation}\label{AdS Fermion Action}
\begin{split}
\mathds{S}_{E} = - \int_{\mathcal{M}_{E}} \mathrm{d}^{d+1}{x} \  \sqrt{g} \ \overbar{\Psi } \, \Big(\Gamma ^{A}  \nabla_{A} - m \Big) \Psi + \mathds{S}_{\partial \mathcal{M}_{E}} \, , 
\end{split}
\end{equation}
where $\overbar{\Psi } = \Psi ^{\dagger} \, \Gamma ^{(v_{E})}$,  
\begin{equation} 
\begin{split}
\nabla _{A} = \partial _{A} + \frac{1}{4} \, \omega _{ab \, A} \, \Gamma ^{ab}.
\end{split}
\end{equation}
where the $\omega _{ab \, A}$ are, in turn, the spin connection 1--forms. The action \eqref{AdS Fermion Action} is the sum of two terms. The first is a bulk term that determines the equations of motion. The second is a boundary term that does not alter the equations of motion. It is fixed by demanding a well--posed variational principle for this system, and will play a crucial role in what is to follow. More will be said about it soon.

The AdS/CFT correspondence posits an equivalence between a dynamical theory of gravity on AdS$_{d+1}$ (the bulk) and a $d$--dimensional CFT that lives on the boundary of this spacetime. In this instance, the bulk fermionic field $\Psi $ is mapped to a fermionic operator $\mathcal{O}$ in the conformal field theory. The precise statement of the duality relates the on--shell bulk action for the field $\Psi $ with the generating functional for connected correlation functions of $\mathcal{O}$.
\begin{equation}\label{AdS CFT fermions}
\begin{split}
\Bigg\langle \exp \Big[\int \mathrm{d}^{d}{x} \ \big(\overbar{\psi } \, \mathcal{O} + \overbar{\mathcal{O}} \, \psi  \big)\Big] \Bigg\rangle_{\mathrm{CFT}} = e^{-\mathds{S}_{E}^{\mathrm{cl}}[\psi ,  \, \overline{\psi } ]}
\end{split}
\end{equation}
Here $ \, \mathds{S}_{E}^{\mathrm{cl}}[\psi , \,  \overbar{\psi }] \, $ is the on--shell action, which is computed by evaluating \eqref{AdS Fermion Action} on a classical solution $\Psi $. The arguments of the on--shell action lets us know how this solution is selected: in addition to being regular in the interior, $\Psi $ must satisfy a suitable Dirichlet boundary condition set by the boundary spinor $\psi $ at $r \to \infty$. It is the interpretation of this boundary condition that introduces two subtleties for fermions. First, the spinors $\Psi $ and $\psi $ belong to spacetimes with different dimensions, and therefore may have a different number of components. Secondly, the Dirac equation that \eqref{AdS Fermion Action} gives rise to is a first order differential equation; na\"{i}vely imposing a Dirichlet boundary condition on a solution to this equation would fix it everywhere without guaranteeing its regularity in the interior. Of course, these two issues are closely related to one another and have been carefully examined and elucidated before, most notably in \cite{Iqbal:2009fd}. Here, we quickly summarize their prescription, directing the reader to the extremely readable original for further details. 

As a preliminary step, it is useful to first split the spinor $\Psi $ in terms of its eigenvalues under the $\Gamma ^{(r)}$ matrix.
\begin{equation} 
\begin{split}
\Psi  = \Psi _{+} + \Psi _{-} \, , \qquad \qquad \Psi _{\pm} = \mathcal{P}_{\pm}  \Psi .
\end{split}
\end{equation}
Using this decomposition in \eqref{AdS Fermion Action}, one immediately sees that $\Psi _{\pm}$ are conjugate to one another. The resolution is to demand the regularity of the solution $\Psi $ in the interior of the bulk geometry --- this naturally allows us to apply Dirichlet boundary conditions on only half the components of $\Psi $, i.e., either on $\Psi _{+}$ or $\Psi _{-}$. 

This prescription neatly resolves the issue of $\Psi $ and $\psi $ having (in general) a different number of components. Recall that the bulk spinor $\Psi $ is a Dirac spinor \eqref{AdS Fermion Action}. When the dimension of the bulk geometry is odd $(d+1 = 2n+1)$, we are free to decompose the \emph{boundary} Dirac spinors into Weyl spinors, which transform under the $(\Sigma , \overbar{\Sigma })$ matrices. Correspondingly, we choose the bulk Gamma  matrices\footnote{The conventions that we adopt differ slightly from those used in \cite{Iqbal:2009fd}. Appendix \ref{Spinor and Gamma Conventions} contains a detailed discussion.} as follows,
\begin{equation} \label{odd d}
\begin{split}
&\Gamma ^{(\mu )} = \begin{pmatrix}
0 & \Sigma ^{(\mu )} \\
- \overbar{\Sigma }^{(\mu )} & 0
\end{pmatrix} , \qquad \Gamma ^{(r)} = \begin{pmatrix}
\mathds{1} & 0 \\
0 & - \mathds{1}
\end{pmatrix}.
\end{split}
\end{equation} 
Importantly, the radial Gamma matrix $\Gamma ^{(r)}$ and the chirality matrix of the boundary theory coincide; with this identification the probe spinor $\psi $ transforms as a $d$--dimensional Weyl spinor. Thus when $ \, d +1= 2n + 1 \, $, the bulk Dirac spinor $\Psi $ is mapped to boundary Weyl spinor primary $\mathcal{O}$.

On the other hand, when the dimension of the bulk is even $d + 1 = 2n$, we make the following choice for our Gamma  matrices,
\begin{equation} \label{even d}
\begin{split}
&\Gamma ^{(\mu )} = \begin{pmatrix}
0 & \gamma  ^{(\mu )} \\
\gamma ^{(\mu )} & 0
\end{pmatrix} , \qquad \Gamma ^{(r)} = \begin{pmatrix}
\mathds{1} & 0 \\
0 & - \mathds{1}
\end{pmatrix}.
\end{split}
\end{equation} 
In this case, the boundary spinor $\psi $ transforms as a Dirac spinor. Thus, when $d+1 = 2n$, the bulk Dirac spinor $\Psi $ is mapped to a boundary Dirac spinor $\mathcal{O}$. In all cases then, the number of components of the bulk Dirac spinor $\Psi $ is always double that of the the boundary spinor $\psi $ and the operator $\mathcal{O}$.

To figure out which to half of the spinor $\Psi $ to provide Dirichlet boundary conditions, we have to look at its asymptotic fall--offs near the AdS boundary. Solving the Dirac equation arising from \eqref{AdS Fermion Action}, we have,
\begin{equation}\label{Fall-offs}
\begin{split}
&\lim_{r \rightarrow \infty} \Psi _{+}(r,k) = \mathsf{A}_{+}(k) \, r^{-\frac{d}{2} + m} + \mathsf{B}_{+}(k) \, r^{-\frac{d}{2} - m - 1} 
\\ 
&\lim_{r \rightarrow \infty} \Psi _{-}(r,k) =  \mathsf{A}_{-}(k) \, r^{-\frac{d}{2} - m} + \mathsf{B}_{-}(k) \, r^{-\frac{d}{2} + m -1} 
\end{split}
\end{equation}
Clearly, the transformation $m \longleftrightarrow -m$ exchanges $\Psi _{+}$ and $\Psi _{-}$, with $\mathsf{A}_{+} \longleftrightarrow \mathsf{A}_{-}$ and $\mathsf{B}_{+} \longleftrightarrow \mathsf{B}_{-}$. In what follows, we will choose $m\geq 0$. This choice selects the fall-off with coefficient $\mathsf{A}_{+}$ as the dominant contribution across all terms in \eqref{Fall-offs} and should therefore be set equal to the source $\psi (k)$. This fixes the Dirichlet condition for $\Psi $ that we seek, to wit, 
\begin{equation}\label{Dirichlet condition}
\begin{split}
\mathsf{A}_{+}(k) = \mathcal{P}_{+} \, S_{0} \,  \psi (k) \, , \qquad \quad \lim_{r \rightarrow \infty} \, r^{\frac{d}{2} - m} \, \Psi _{+} =  \mathcal{P}_{+} \, S_{0} \,  \psi .
\end{split}
\end{equation}
Here, $S_{0}$ is a constant matrix that is defined in Appendix \ref{Spinor and Gamma Conventions}. As a result, the conformal dimension $\Delta $ of the boundary operator $\mathcal{O}$ is,
\begin{equation} \label{Conformal dimension}
\begin{split}
\Delta = \frac{d}{2} + m.
\end{split}
\end{equation}

Knowing the Dirichlet condition for the bulk fermion $\Psi $, in turn, determines the boundary term in the action \eqref{AdS Fermion Action}. Having chosen to fix $\Psi _{+}$ at the AdS boundary, we are no longer free to vary $\Psi _{-}$. As a result, the variation of $\mathds{S}_{E}$ shouldn't contain any terms proportional to $\delta \Psi _{-}$. This can be arranged by choosing $\mathds{S}_{\partial \mathcal{M}_{E}}$ to be,
\begin{equation}\label{Dirac action boundary term}
\begin{split}
\mathds{S}_{\partial \mathcal{M}_{E}}= - \int_{\partial \mathcal{M}_{E}} \mathrm{d}^{d}{x} \ r^{d} \ \overbar{\Psi } \; \mathcal{P}_{-} \Psi .
\end{split}
\end{equation}

We can now place the Dirac action \eqref{AdS Fermion Action} on--shell. Notice that the entire contribution arises from the boundary term \eqref{Dirac action boundary term}; the first term in \eqref{AdS Fermion Action} vanishes when evaluated on a solution of the EOM. However, as can be seen from $\eqref{Fall-offs}$, this quantity has divergent contributions if $m > 1/2$. The standard holographic prescription asks us to select only the finite pieces from the on--shell action, which is the end result of a careful holographic renormalization \cite{Ammon:2010pg}, i.e., 
\begin{equation}\label{}
\begin{split}
\mathds{S}_{E}^{\mathrm{ren}}[\psi , \,  \overbar{\psi }] = \Big\{\mathds{S}_{E}^{\mathrm{cl}}[\psi ,  \, \overbar{\psi }]\Big\}_{r^{0}} = - \int_{\partial \mathcal{M}_{E}} \mathrm{d}^{d}{x} \ r^{d} \ \overbar{\mathsf{A}}_{+} \, \mathsf{A}_{-} .
\end{split}
\end{equation}
Using \eqref{AdS CFT fermions} we can now compute \emph{renormalized} Euclidean correlation functions for $\mathcal{O}$ and $\overbar{\mathcal{O}}$. Of course, in order to generate retarded and advanced correlators, we would have to analytically continue these results to real--time.

In order to generate Schwinger--Keldysh correlators we will upgrade the above discussion to an appropriately `doubled' version of an asymptotically AdS spacetime. While detailing this prescription in $\S$\ref{Subsection: Open EFT from Holography}, we will work directly in a Lorentzian spacetime, using the real--time versions of the expressions presented here. These can be obtained by the following analytic continuation,
\begin{equation}\label{Wick rotation}
\begin{split}
v_{E} \to i \, v \, , \qquad \delta _{ab} \to \eta _{ab} \, , \qquad \Gamma ^{(v_{E})} \to i \, \Gamma ^{(v)} \, , \qquad  \mathds{S}_{E} \to -i \,  \mathds{S}.
\end{split}
\end{equation}

\subsection{The gravitational Schwinger--Keldysh saddle}\label{Subsection: grSK geometry}
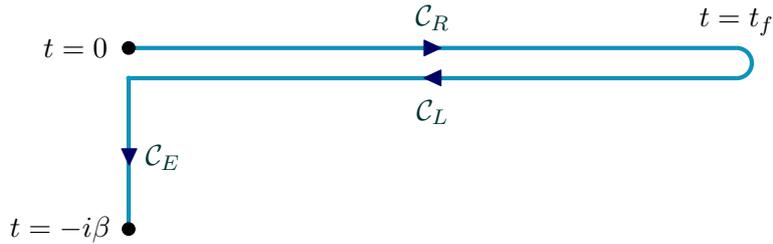
\begin{figure}
	\begin{center}
		\begin{tikzpicture}[]
		
		\draw[ultra thick, cyan!40!teal] (0,0) -- (8,0);
		\draw[ultra thick, cyan!40!teal] (8,-0.4) -- (0,-0.4);
		\draw[ultra thick, cyan!40!teal] (0,-0.4) -- (0,-2.4);
		\draw[ultra thick, cyan!40!teal] (8,0) arc (90: -90 : 0.2);
		
		\node[circle, fill, black, scale= 0.5] at (0,0) {};
		\node[circle, fill, black, scale= 0.5] at (0,-2.4) {};
		\node[circle, fill, cyan!60!teal, scale= 0.18] at (0,-0.4) {};
		
		\node[blue!40!black] at (4,0) {$\blacktriangleright$};
		\node[blue!40!black] at (4,-0.4) {$\blacktriangleleft$};
		\node[blue!40!black] at (0.01,-1.45) {$\blacktriangledown$};
		
		\node at (8,0.35) {$t = t_{f}$};
		\node at (-0.7,0) {$t=0$};
		\node at (-.9,-2.4) {$t = -i \beta$};
		\node[text = teal!40!black] at (0.45,-1.45) {$\mathcal{C}_{E}$};
		\node[text = teal!40!black] at (4,0.4) {$\mathcal{C}_R$};
		\node[text = teal!40!black] at (4,-.85) {$\mathcal{C}_L$};
		\end{tikzpicture}
	\end{center}
	\caption{\label{fig:fixedRho} The Schwinger--Keldysh complex time contour. The direction of the contour is as indicated, with the forward and backward evolutions occuring along the sections $\mathcal{C}_R$ and $\mathcal{C}_L$, respectively. The length of the Euclidean time segment, $\mathcal{C}_{E}$, is determined by the inverse temperature $\beta $.}
\end{figure}

As we had described in $\S$\ref{Fermions in SK}, the Schwinger--Keldysh contour for a thermal state is a \emph{complex} time path, travelling from $t = 0$ to $t = T$, looping back to $t = 0$ and proceeding to $t = - i \beta $, as illustrated in Fig.\ref{fig:fixedRho}. In order to have a holographic interpretation of this, we would like to supplement the contour by a suitable bulk geometry, in a manner that the contour forms a time--like codimension--1 boundary of the larger spacetime.

We begin by writing the asymptotically AdS$_{d+1}$ black brane in ingoing Eddington--Finkelstein coordinates.
\begin{equation}\label{AdS BB Metric}
\begin{split}
\mathrm{d}s^{2} &= - r^{2} \, f(r) \, \mathrm{d}v^{2} + 2 \, \mathrm{d}v \, \mathrm{d}r  + r^{2} \, \mathrm{d}\bm{x} ^{2}_{d-1}, \qquad \qquad f(r) = 1 - \frac{r_{h}^{d}}{r^{d}}
\end{split}
\end{equation}
Here $r_{h}$ is the radius of the horizon and the AdS radius has been set to unity. The temperature of this black brane is given by,
\begin{equation}\label{Inverse temperature}
\begin{split}
\beta = \frac{4 \pi }{d \,  r_{h}}
\end{split}
\end{equation}

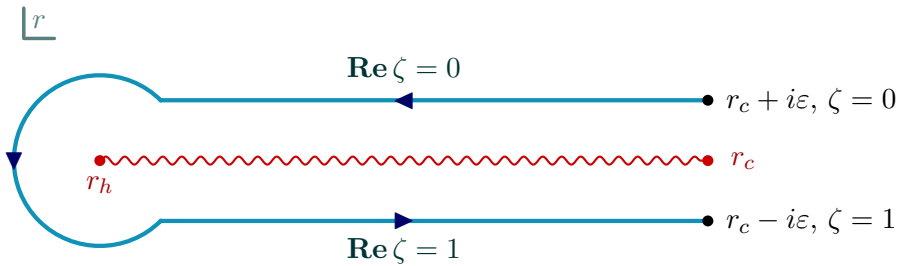
\begin{figure}
	\begin{center}
		\begin{tikzpicture}[scale=0.8]
		
		\node[teal!30!gray, scale=3] at (-1, 2.3) {\(\llcorner \)} ;
		\node[teal!30!gray, scale=1] at (-1, 2.3) {\(r \)} ;
		
		\draw[cyan!40!teal, ultra thick] (1,1) -- (10,1);
		\draw[cyan!40!teal, ultra thick] (1,1) arc (45: 315 : 1.414);
		\draw[cyan!40!teal, ultra thick] (1,-1) --(10,-1);
		\draw[decoration={snake, segment length = 2.5 mm, amplitude=1.5}, decorate, red!80!black, thick] (0,0) -- (10,0);
		
		\node[circle,scale= 0.4,red!80!black,fill] at (0,0) {} ;
		\node[circle,scale= 0.4,red!80!black,fill] at (10,0) {} ;
		\node[circle, fill, cyan!40!teal, scale= 0.17] at (1,1) {} ;
		\node[circle, fill, cyan!40!teal, scale= 0.17] at (1,-1) {};
		\node[circle, fill, black, scale= 0.4] at (10,1) {} ;
		\node[circle, fill, black, scale= 0.4] at (10,-1) {} ;
		
		\node[blue!40!black] at (5,1) {$\blacktriangleleft$};
		\node[blue!40!black] at (5,-1) {$\blacktriangleright$};
		\node[blue!40!black] at (-1.41,0) {$\blacktriangledown$};

		\node[teal!40!black] at (5,1.5) {$\mathbf{Re} \,  \zeta = 0$};
		\node[teal!40!black] at (5,-1.5) {$\mathbf{Re} \, \zeta  = 1$};
		\node[red!65!black] at (0,-0.4) {$r_h$};
		\node[red!65!black] at (10.58,0) {$r_c$};
		\node at (11.7,1) {$r_{c}+i\varepsilon, \, \Ctor=0$};
		\node at (11.7,-1) {$r_{c}-i\varepsilon, \, \Ctor=1$};

		\end{tikzpicture}
		\caption{\label{fig:fixedv} The radial contour drawn on the complex $r$ plane, at fixed $v$. The locations of the two boundaries and the horizon have been indicated, along with the direction of the contour. The red line denotes the branch cut beginning at the horizon.}
	\end{center} 
\end{figure}

In order to realize the Schwinger--Keldysh contour on the boundary theory, we promote the coordinate $r$ to a complex variable and select a suitable line in the complex plane as our radial coordinate. The  particular curve we choose is the following: we select two copies of a line extending from the black hole horizon to infinity, each infinitesimally displaced in the imaginary direction on either side of the real line. That is, one line extends from $r= r_{h} + i \epsilon $ to $r = \infty + i \epsilon $ and the other extends from $r = r_{h} - i \epsilon $ to $r = \infty - i \epsilon $. We then join these two legs by a \emph{horizon cap} -- a smooth curve connecting the two points $r = r_{h} \pm i \epsilon $, encircling the point $r = r_{h}$.  For computational ease, it will prove convenient to terminate both these legs at finite cut-offs, $r = r_{c} \pm i\epsilon $. We will eventually take $r_{c} \rightarrow  \infty$ at the end of our calculations. This is the radial contour for the full spacetime and is illustrated in Fig.\ref{fig:fixedv}.

We will find it useful to utilise a different parametrization for the radial direction, dubbed the \emph{mock tortoise} coordinate $\zeta $. This is defined by the differential equation,
\begin{equation}\label{Mock Tortoise}
\begin{split}
\frac{\mathrm{d}{r}}{\mathrm{d}{\zeta }} = \frac{i \beta }{2} \, r^{2}  f(r)
\end{split}
\end{equation}
This coordinate is not an analytic function of $r$ on the entire complex plane, and instead has a logarithmic branch point at $r = r_{h}$, the zero of the emblackening factor $f(r)$ -- we choose the corresponding branch cut to traverse from $r = r_{h}$ to $r = \infty$. With our choice of normalization, the monodromy that $\zeta $ picks up around this branch is unity. 

Evaluated on each leg of the radial contour described above, the mock tortoise coordinate is a complex parameter, with an imaginary part running from $0$ at the AdS boundary to $\infty$ at the horizon. The real parts of the two legs differ by the monodromy around the horizon, viz., unity. As a matter of convention, we choose its real part to vanish on the upper leg of the contour, and set it to unity on the lower leg -- this can now be used to distinguish between the two. Thus, the mock tortoise coordinate asymptotes to,
\begin{equation}\label{Finite cut-off}
\begin{split}
\zeta (r_{c} + i\epsilon ) = 0, \qquad \qquad \zeta (r_{c}  - i \epsilon ) = 1
\end{split}
\end{equation}
Indeed, the mock tortoise coordinate $\zeta $ can be immediately integrated in terms of $r$ to give,
\begin{equation} 
\begin{split}
\zeta + \zeta _{c} = \frac{i}{2 \pi } \,  \frac{d}{d-1} \, \left(\frac{r_{h}}{r}\right)^{d-1}  {} _{2}F_{1} \left(1, 1- \frac{1}{d} ; 2 - \frac{1}{d} ; \frac{r^{d}}{r_{h}^{d}}\right)
\end{split}
\end{equation}
Here $\zeta _{c}$ is a constant chosen to set $\zeta (r_{c} + i\epsilon ) = 0$. A constant time slice of the mock tortoise co-ordinate in the complex plane, along with our choice of branch cut is depicted in Fig.\ref{fig:fixedv}.

\begin{figure}
	\centering
	\subfloat{\includegraphics[width=0.25\textwidth]{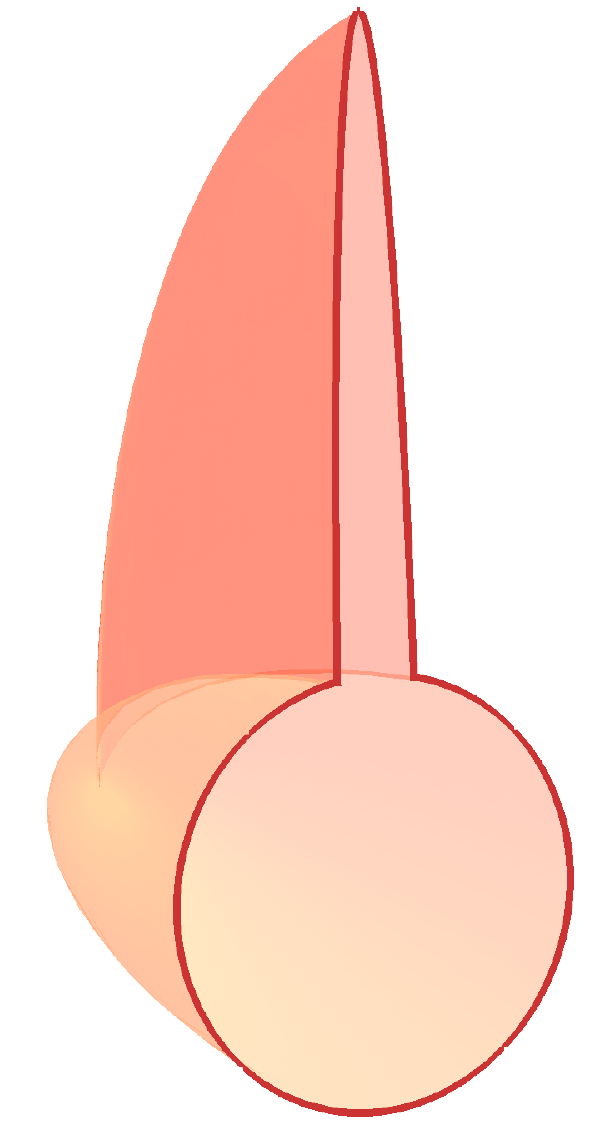}}
	\qquad \qquad \qquad 
	\subfloat{\includegraphics[width=0.35\textwidth]{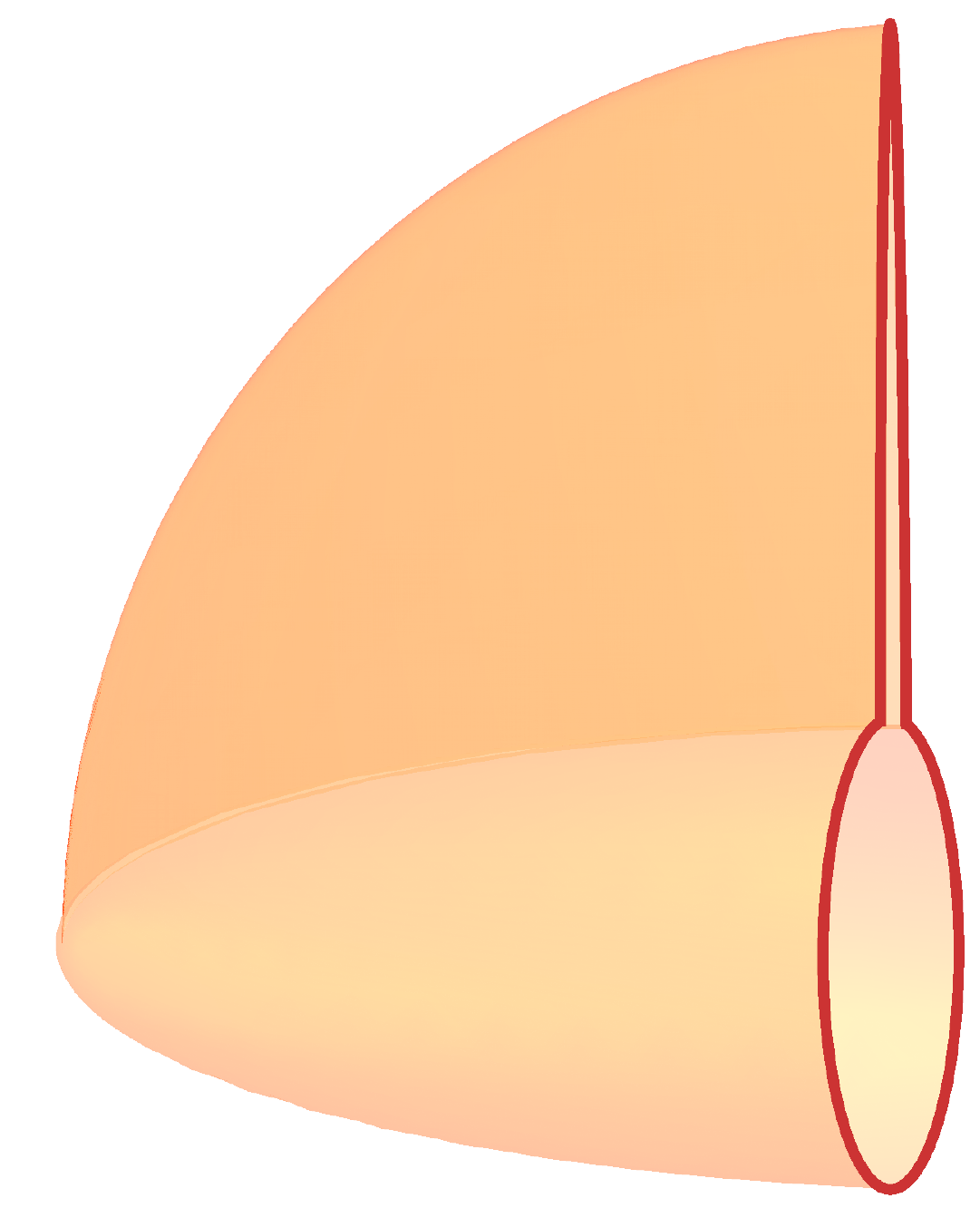}}
	\caption{\label{fig:SKfig} Two perspectives of the complex two--sheeted grSK geometry. The left panel
		displays the boundary thermal SK contour which is filled in the Euclidean portion by the Euclidean black hole (cigar)
		geometry and in the Lorentzian section by two copies of the exterior of a Lorentzian black hole spacetime. The right panel emphasize the smooth join of
		the two sheets of the Lorentzian section. (The above figure is from \cite{Jana:2020vyx}.)}
\end{figure}

When expressed in these coordinates, the metric reads,
\begin{equation}\label{Metric}
\begin{split}
\mathrm{d}s^{2} = - r^{2} \, f \, \mathrm{d}v^{2} + i \, \beta \,  r^{2} \,  f \, \mathrm{d}\zeta \, \mathrm{d}v + r^{2} \, \mathrm{d}\bm{x} ^{2}_{d-1} .
\end{split}
\end{equation}
The spacetime thus consists of two manifolds $\mathcal{M}_{L}$ and $\mathcal{M}_{R}$ (with $\mathbf{Re} \, \zeta =0, \, 1$  respectively), smoothly joined along their radial directions by the horizon cap. The advanced time $v$ is now identified with the time coordinate $t$ on the boundary of this spacetime at $r \rightarrow \infty \pm i\epsilon$. In order to make contact with the Schwinger-- Keldysh contour, these two manifolds require one further identification -- each radially constant slice of $\mathcal{M}_{L}$ and $\mathcal{M}_{R}$ meets at the future turning point, $v \rightarrow \infty$. Thus, the sheets $\mathcal{M}_{L}$ and $\mathcal{M}_{R}$ are the bulk extensions of the $\mathcal{C}_{L}$ and $\mathcal{C}_{R}$ regions of the Schwinger--Keldysh contour, respectively. This resulting geometry is called the \emph{gravitational Schwinger--Keldysh saddle} \cite{Jana:2020vyx}, and is illustrated in Fig.\ref{fig:SKfig}.

\subsection*{A grSK involution isometry} 
To conclude our discussion of the geometry, we record a feature that will play an important role in the analysis to follow. The gravitational Schwinger--Keldysh saddle is  invariant under the following involutive diffeomorphism,
\begin{equation}\label{Involutive diffeomorphism}
\begin{split}
v \longmapsto i\beta \zeta - v, \qquad \bm{x} \longmapsto - \bm{x}.
\end{split}
\end{equation}
It is immediately clear that the metric \eqref{Metric} is unchanged by this transformation. While it is tempting to interpret this isometry as a $PT$ reversal of the coordinates on the boundary of the spacetime, this isn't quite correct: the transformation of the advanced Eddington--Finkelstein time $v$ depends on the radial coordinate, $\zeta $. Indeed, these geometries do not possess a time reversal symmetry, as we may have anticipated for a gravitational dual to a non-equilibrium field theory.  As noted by the authors of  \cite{Chakrabarty:2019aeu, Jana:2020vyx}, the above symmetry\footnote{In  \cite{Chakrabarty:2019aeu,Jana:2020vyx}, the authors do not flip the spatial co-ordinates $\bm{x}$.} is extremely useful for obtaining propagating solutions. In the following sections, we will exploit this involution extensively to construct Green's functions that solve the Dirac equation in the grSK geometry.

\subsection{Open EFTs from holography}\label{Subsection: Open EFT from Holography}
In order to derive the Schwinger--Keldysh generating functional for the operator $\mathcal{O}$, all that is left to do now is to lift the prescription in $\S$\ref{Subsection: Fermions in AdS/CFT} to the grSK saddle. This is easily done, switching coordinates, and writing the Lorentzian version of the action \eqref{AdS Fermion Action} as a contour integral over the doubled geometry.
\begin{equation}\label{grSK Fermion Action}
\begin{split}
\mathds{S}_{\Psi } =  \ointctrclockwise \mathrm{d}{\zeta }  \int \mathrm{d}^{d}{x} \ \sqrt{-g} \,  i \, \overbar{\Psi } \, \Big(\Gamma ^{A } D_{A } - m \Big) \Psi  +  \int \mathrm{d}^{d}{x} \ r^{d} \ i \,  \overbar{\Psi } \, \ProCtor_{-} \Psi \  \Bigg|^{\zeta = 1}_{\zeta = 0}
\end{split}
\end{equation}
Notice that the boundary term is now evaluated at both the right and the left boundaries of the grSK spacetime. The rest of the recipe proceeds analogously: we first need to solve for the fermion $\Psi $ in this spacetime. Crucially, we will now impose Dirichlet boundary conditions at \emph{both} the (right and left) boundaries of the grSK saddle,
\begin{equation}\label{grSK Dirichlet conditions}
\begin{split}
\lim_{\zeta \to 0} \ r^{\frac{d}{2} - m} \, \mathcal{P}^{\zeta }_{+} \Psi = S_{0}\psi _{L} \, , \qquad \qquad \lim_{\zeta \to 1} \ r^{\frac{d}{2} - m} \, \mathcal{P}^{\zeta }_{+} \Psi = S_{0}\psi _{R}.
\end{split}
\end{equation}
In contrast to the situation in $\S$\ref{Subsection: Fermions in AdS/CFT}, such a solution will (in general) \emph{not} be regular in the interior of the grSK geometry, with branch--cuts at the horizon of the AdS black brane. As we will see, these branch--cuts encode a lot of the physics of these systems and have to be dealt with carefully. Having obtained a solution with these specifications, we can now evaluate the action \eqref{grSK Fermion Action} on--shell, again selecting only its finite part to give us the generating functional for renormalized Schwinger--Keldysh correlators of the boundary CFT, $\mathcal{Z}_{\mathrm{SK}} \, [\psi _{R}, \, \overbar{\psi }_{R}, \, \psi _{L} , \, \overbar{\psi }_{L}]$. Deriving the above generating functional is equivalent to deriving the open effective field theory that we seek.

\section{Classical origin of Fermi--Dirac statistics}\label{sec:FermionicProp}
Having outlined how to derive an open effective field theory using holographic methods, the next task at hand is to study the dynamics of a fermionic field propagating in the bulk geometry. As emphasized in the previous section the influence phase for the probe field is obtained by taking the Dirac action on--shell --- to do so, we have to solve the Dirac equation in this spacetime, subject to the boundary conditions \eqref{grSK Dirichlet conditions}. This is what we turn to now.

We organize this section as follows: we begin by introducing the Dirac equation for the bulk fermion in $\S$\ref{Subsection: The Dirac Equation}. In $\S$\ref{Subsection: Time Reversing Solutions}, we then detail how to use the involutive diffeomorphism \eqref{Involutive diffeomorphism} to `time--reverse' solutions to this wave equation, allowing us to generate a full solution once the solution satisfying the infalling boundary conditions is known. As an important consequence, in $\S$\ref{Subsection: Fermi Dirac statistics} we will see how this leads to the correct statistics for this system. Once these constructions have been laid out, we will move on to computing the infalling boundary--to--bulk Green's function. In general $d$, the Dirac wave equation is not known to admit closed form solutions in these spacetimes. However, we can solve them in a gradient expansion, i.e., determining them order by order in increasing powers of frequency and momenta. This is carried out in $\S$\ref{sec:GradientExp}.

\subsection{The Dirac equation.}\label{Subsection: The Dirac Equation}
In order to set up the Dirac equation in this geometry, we use the following tangent frame \cite{Ceplak:2019ymw},
\begin{equation}\label{Tetrads}
\begin{split}
&E^{(v)} = \frac{r}{2}   \, \mathrm{d}v -  \frac{ f}{2} \, r \, \big(i \beta \, \mathrm{d}\zeta - \mathrm{d}v \big), \qquad E^{(\zeta )}  = \frac{r}{2}   \, \mathrm{d}v +  \frac{ f}{2} \, r \, \big(i \beta \, \mathrm{d}\zeta - \mathrm{d}v \big), \qquad E^{(i)} = r \, \mathrm{d}x^{i} \, . 
\end{split}
\end{equation}
These tetrads have two advantages. First, none of the components of the frame or their derivates are badly behaved at the blackhole horizon. Second, there are no square roots of the the emblackening factor $f$ that appear. These features are of particular importance to the grSK saddle because it involves an analytical continuation of the radial direction across the future horizon --- the above tangent frame is thus smooth on the whole spacetime. The spin connection 1--forms corresponding to these tetrads are,
\begin{equation}\label{Spin connection 1 forms}
\begin{split}
\omega _{(v) (\zeta)} = \Big(r f - \frac{1}{i\beta f}\frac{\mathrm{d}f}{\mathrm{d}{\zeta }} \Big) \, \mathrm{d}v - \frac{i\beta}{2} \,  r f \, \mathrm{d}\zeta, \quad \ \omega _{(i)(v)} = - \frac{1 - f}{2} \, r \, \mathrm{d}x^{i}, \quad \ \omega _{(i)(\zeta) } =  \frac{1 + f}{2} \, r \,  \mathrm{d}x^{i},
\end{split}
\end{equation}
with all other components, not related by symmetry, vanishing. A little algebra then leads us to the following Dirac equation,
\begin{equation}\label{Dirac Equation}
\begin{split}
\bigg[ \, \mathbb{\Gamma } \, \big(\partial _{\zeta } + i\beta \, \partial _{v}\big) + \mathbb{\Gamma }^{\dagger} \, f^{-1} \, \partial _{\zeta } &+ \frac{i\beta}{2}  \, \big(\Gamma ^{(i)} \, \partial _{i} - m r\big) 
\\
&+ \frac{1}{2} \ \mathbb{\Gamma } \, \frac{\mathrm{d}}{\mathrm{d}{\zeta }} \log \big(r^{d} f \big) +  \frac{1}{2} \ \mathbb{\Gamma }^{\dagger} \,  f^{-1} \, \frac{\mathrm{d}}{\mathrm{d}{\zeta }} \log (r^{d}) \, \bigg] \Psi  = 0,
\end{split}
\end{equation}
where we have introduced the matrices $\mathbb{\Gamma }$ and $\mathbb{\Gamma }^{\dagger}$ to keep expressions compact.
\begin{equation}\label{BBGamma}
\begin{split}
\mathbb{\Gamma } = \frac{1}{2} \, \big(\Gamma ^{(\zeta )} + \Gamma ^{(v)}\big) \ \ \implies \ \ \mathbb{\Gamma }^{\dagger} = \frac{1}{2} \, \big(\Gamma ^{(\zeta )} - \Gamma ^{(v)}\big) \, .
\end{split}
\end{equation}
We emphasize that by definition \eqref{Mock Tortoise}, both $\partial _{\zeta }$ and $f^{-1} \, \partial _{\zeta }$ are regular vector fields at the future horizon. Consequently, despite initial appearances, both the spin connection 1--forms and the Dirac equation are well defined along the holographic contour. 

As a final observation, we note that the metric components \eqref{Metric} are independent of $v$ and $\bm{x}$. Thus, it is useful to perform a Fourier transformation with respect to these coordinates, allowing these modes to decouple. We adopt the following shorthand for the Fourier transform,
\begin{equation}\label{Fourier transform}
\begin{split}
\Psi (v,\zeta , \bm{x}) = \int \frac{\mathrm{d}{\omega } \ \mathrm{d}^{d-1}{\bm{k}}}{(2 \pi )^{d}} \ \Psi (\omega , \zeta , \bm{k}) \, e^{- i \omega v + i \bm{k} \cdot \bm{x}} \equiv \int_{k}  \Psi(\omega ,\zeta , \bm{k}) \, e^{ikx}.
\end{split}
\end{equation}
The field $\Psi (\omega ,\zeta ,\bm{k})$ is then easily seen to satisfy,
\begin{equation}\label{Dirac Equation FT}
\begin{split}
\bigg[ \, \mathbb{\Gamma } \, \big(\partial _{\zeta } + \beta \omega \big) + \mathbb{\Gamma }^{\dagger} \, f^{-1} \, \partial _{\zeta } &- \frac{\beta}{2}  \, \big(\Gamma ^{(i)} \, k_{i} + i m r\big) 
\\
&+ \frac{1}{2} \ \mathbb{\Gamma } \, \frac{\mathrm{d}}{\mathrm{d}{\zeta }} \log \big(r^{d} f \big) +  \frac{1}{2} \ \mathbb{\Gamma }^{\dagger} \,  f^{-1} \, \frac{\mathrm{d}}{\mathrm{d}{\zeta }} \log (r^{d}) \, \bigg] \Psi  = 0.
\end{split}
\end{equation}

\subsection{`Time--reversing' solutions.}\label{Subsection: Time Reversing Solutions} 
In the following sections, we will solve the above Dirac equation \eqref{Dirac Equation FT} for the boundary--to--bulk Green's function. As we will see, these propagators come in two types: an \emph{infalling} (\emph{retarded}) propagator and an \emph{outgoing} (\emph{advanced}) one\footnote{The terms infalling and outgoing are used in reference to the radial direction of the geometry.}, with the full boundary--to--bulk Green's function being a linear combination of the two. The involutive diffeomorphism described at the end of $\S$\ref{Subsection: grSK geometry} allows us to exploit this structure: using \eqref{Involutive diffeomorphism}, we can construct a map between infalling and outgoing solutions. Thus, once the infalling Green's function is known, the full boundary--to--bulk Green's function can be immediately written down.

This construction isn't new --- the utility of this symmetry for this very purpose was first recognized and exploited by the authors of \cite{Chakrabarty:2019aeu} in the context of a classical string probing the grSK geometry, and later in \cite{Jana:2020vyx} for a probe scalar field. In this section, we describe how this construction works for the Dirac field. As we will see, this discussion contains subtleties not present in the other two cases.
 
 To begin with, let us consider the effect of the global coordinate transformation (GCT) \eqref{Involutive diffeomorphism} on the Dirac operator \eqref{Dirac Equation} --- it simply exchanges the derivatives $\partial _{\zeta } \longleftrightarrow \partial _{\zeta } + i \beta \, \partial _{v}$, and therefore does \emph{not} leave the operator unchanged. This is to be expected --- crucially, the form of the Dirac equation depends both on our choice of coordinates and our choice of tangent frame. While the isometry leaves the metric invariant, it has a non--trivial action on the space of 1--forms induced by its Jacobian $\mathcal{J}^{A}_{\ \ B}$, 
\begin{equation} 
\begin{split}
E^{a}_{A} \longrightarrow  \mathcal{J}_{A}^{\  B} \; E^{a}_{B}  \, ,  \qquad \mathrm{where} \ \ \quad \mathcal{J}_{A}^{\  B} = \begin{pmatrix}
-1 & 0 & 0 \\
i \beta & 1  & 0 \\
0 & 0 & -\mathds{1}
\end{pmatrix}.
\end{split}
\end{equation}
 
In order to realize the involution \eqref{Involutive diffeomorphism} as a discrete symmetry of the Dirac equation, we have to compensate for the above transformation of the tetrads by an appropriate linear transformation of the tangent frame. In other words, we seek a matrix $\mathcal{T}^{a}_{\ \ b}$ that satisfies,
\begin{equation}\label{Tetrads Symmetry}
\begin{split}
\mathcal{T}^{a}_{\ \ b}  \ \mathcal{J}_{A}^{\  B} E^{b}_{B} \;   = E^{a}_{A}.
\end{split}
\end{equation}
This matrix equation is easily solved, giving,
\begin{equation}\label{PT LLT}
\begin{split}
\mathcal{T}^{a}_{\ \ b} = \begin{pmatrix}
-1 & 0 & 0 \\
0 & 1 & 0 \\
0 & 0 & -\mathds{1}
\end{pmatrix}\begin{pmatrix}
\cosh \vartheta  & \ \sinh \vartheta  & \ 0
\\
\sinh \vartheta  & \ \cosh \vartheta  & \ 0 
\\
0 & \ 0 & \ \mathds{1}
\end{pmatrix}, \quad \vartheta = \log f.
\end{split}
\end{equation}
Thus, the linear transformation $\mathcal{T}$ is actually an idempotent local Lorentz transformation (LLT) composed of a boost along $(\zeta )$ with rapidity parameter $\vartheta$ followed by a reflection of the $(v)$ and $(\bm{x})$ axes. This LLT is an improper transformation when the dimension of the bulk is even, and proper when its dimension is odd. 

With the metric and the tetrads left invariant, the involution \eqref{Involutive diffeomorphism} can now be regarded as a symmetry of the Dirac operator \eqref{Dirac Equation} --- the lesson from the above calculation is that this requires both the GCT as well a compensatory LLT.


\subsection*{Constructing outgoing solutions.}
We are now in a position to construct the map between infalling and outgoing solutions. Consider an infalling solution of \eqref{Dirac Equation FT}, $\Psi ^{\mathrm{in}}(\omega ,\zeta ,\bm{k})$. Drawing from our previous discussion, in order to generate an outgoing solution from this, simply performing the coordinate transformation \eqref{Involutive diffeomorphism} is not enough --- the involution induces a linear transformation on spinor space that we have to undo by a compensatory transformation, $\mathfrak{T}$. Let us deal with these two steps one by one. The effect of the GCT is given by,
\begin{equation} 
\begin{split}
\Psi ^{\mathrm{in}}(\omega ,\zeta ,\bm{k})  \ \longmapsto \ \Psi ^{\mathrm{in}}(-\omega ,\zeta ,-\bm{k}) \, e^{-\beta \omega \zeta } \equiv \Psi ^{\mathrm{rev}}(\omega ,\zeta ,\bm{k}) \, e^{-\beta \omega \zeta },
\end{split}
\end{equation}
where we have defined the \emph{frequency reversed} spinor, $\Psi ^{\mathrm{rev}}(\omega ,\zeta ,\bm{k})$. The relevant linear transformation in spinor space, $\mathfrak{T}$, is simply the LLT in spinor space that corresponds to the Lorentz transformation $\mathcal{T}$, to wit,
\begin{equation}\label{Spinor PT LLT}
\begin{split}
\mathfrak{T} = \Gamma ^{(\zeta )} \cdot \exp\Big(\, \frac{\vartheta }{2} \ \Gamma ^{(\zeta )} \, \Gamma ^{(v)}\Big) = \sqrt{f} \; \mathbb{\Gamma } + \sqrt{f}^{-1} \, \mathbb{\Gamma }^{\dagger}.
\end{split}
\end{equation}
Indeed, exactly like \eqref{PT LLT}, this matrix is constructed out of two parts. The first factor, $\exp \big(\frac{\vartheta }{2} \, \Gamma ^{(\zeta )} \, \Gamma ^{(v)} \big)$, is a boost in spinor space along $(\zeta )$ with rapidity $\vartheta $. This is then multiplied by $\Gamma ^{(\zeta )}$, a matrix that \emph{anti-commutes} with $\Gamma ^{(v)}$ and all the $\Gamma ^{(i)}$, accounting for the reflection of the $(v)$ and $(\bm{x})$ directions. It is easily check that the matrix $\mathfrak{T}$ is idempotent.

Thus, combining the above, we are led to the following recipe: given an ingoing solution $\Psi ^{\mathrm{in}}(\omega , \zeta , \bm{k})$, we can construct an outgoing solution $\Psi ^{\mathrm{out}}(\omega , \zeta , \bm{k})$ as follows,
\begin{equation}\label{Psi Out}
\begin{split}
\Psi ^{\mathrm{out}}(\omega  ,\zeta , \bm{k}) = \mathfrak{T} \cdot \Psi ^{\mathrm{rev}} (\omega ,  \zeta ,\bm{k}) \, e^{-\beta \omega \zeta } = \mathfrak{T} \cdot \Psi ^{\mathrm{in}} (- \omega ,  \zeta ,- \bm{k}) \, e^{-\beta \omega \zeta }.
\end{split}
\end{equation}
The infalling solution $\Psi ^{\mathrm{in}}(\omega ,\zeta ,\bm{k})$ is (by definition) regular at the future horizon of the black brane. The above construction then implies that the outgoing solution is \emph{not} smooth in this region, with branch--cuts arising from the matrix $\mathfrak{T}$ and the factor $e^{-\beta \omega \zeta }$. This ties in with our previous discussion in $\S$\ref{Subsection: Open EFT from Holography} --- the full solution, which is an appropriate linear combination of the infalling and outgoing solutions is not regular in the interior.

The validity of the above prescription for outgoing solutions is easily checked. Evaluating $\Psi ^{\mathrm{out}}(\omega ,\zeta ,\bm{k})$ on the Dirac operator \eqref{Dirac Equation FT}, a short calculation demonstrates that this spinor solves the wave equation if and only if $\Psi ^{\mathrm{rev}}(\omega ,\zeta ,\bm{k})$ satisfies, 
\begin{equation} 
\begin{split}
\bigg[ \, \mathbb{\Gamma } \, \big(\partial _{\zeta } - \beta \omega \big) + \mathbb{\Gamma }^{\dagger} \, f^{-1} \, \partial _{\zeta } &+ \frac{\beta}{2}  \, \big(\Gamma ^{(i)} \, k_{i} - i m r\big) 
\\
&+ \frac{1}{2} \ \mathbb{\Gamma } \, \frac{\mathrm{d}}{\mathrm{d}{\zeta }} \log \big(r^{d} f \big) +  \frac{1}{2} \ \mathbb{\Gamma }^{\dagger} \,  f^{-1} \, \frac{\mathrm{d}}{\mathrm{d}{\zeta }} \log (r^{d}) \, \bigg] \Psi ^{\mathrm{rev}} = 0.
\end{split}
\end{equation}
This is just the Dirac operator $\eqref{Dirac Equation FT}$ with the signs of its frequency and momenta flipped. By definition, $\Psi ^{\mathrm{in}}(-\omega ,\zeta ,-\bm{k}) = \Psi ^{\mathrm{rev}}(\omega ,\zeta ,\bm{k})$ is a solution to this equation. 



\subsection{The emergence of Fermi--Dirac statistics} \label{Subsection: Fermi Dirac statistics}
Following the prescription described in Section \ref{Subsection: grSK geometry}, we can write the full solution of the Dirac equation on the Schwinger Keldysh contour as,
\begin{equation}\label{Psi full solution}
\begin{split}
\Psi(\zeta , \omega , \bm{k}) =  S^{\mathrm{in}}(\zeta , \omega , \bm{k}) \; c(\omega , \bm{k}) +   \Big[\sqrt{f} \, \mathbb{\Gamma } + \frac{1}{\sqrt{f}} \, \mathbb{\Gamma }^{\dagger} \Big] \; S^{\mathrm{rev}}(\zeta , \omega , \bm{k}) \;  h(\omega , \bm{k}) \; e^{-\beta \omega \zeta } \, .
\end{split}
\end{equation}
Here, $c(\omega, \bm{k})$ and $h(\omega, \bm{k})$ are (\emph{causal} and \emph{Hawking}) boundary spinors. $S^{\mathrm{in}/\mathrm{rev}}(\zeta , \omega , \bm{k})$ are the boundary to bulk Green's functions, and,
\begin{equation}\label{S Rev}
\begin{split}
S^{\mathrm{rev}}(\zeta , \omega , \bm{k}) \equiv S^{\mathrm{in}}(\zeta ,- \omega , - \bm{k}) \, .
\end{split}
\end{equation}

We now impose boundary conditions on this solution using the sources $\psi _{L}(\omega , \bm{k})$ and $\psi _{R}(\omega ,\bm{k})$ at the two conformal boundaries $(r \rightarrow r_{c} \pm i\epsilon )$ of the holographic SK contour.
\begin{equation}\label{psi L R}
\begin{split}
&\lim_{r \rightarrow r_{c} + i\epsilon } r^{\frac{d}{2}-m}\, \mathcal{P}^{\zeta }_{+} \Psi = \mathcal{P}^{\zeta }_{+} \, S_{0} \, \psi _{L} 
\\ 
&\lim_{r \rightarrow r_{c}- i\epsilon } r^{\frac{d}{2}-m}\, \mathcal{P}^{\zeta }_{+} \Psi = \mathcal{P}^{\zeta }_{+} \, S_{0} \, \psi _{R}
\end{split}
\end{equation}
Here, $S_{0}$ is a constant matrix related to $S^{\mathrm{in}/\mathrm{rev}}$ via,
\begin{equation}\label{S0}
\begin{split}
\lim_{r \rightarrow r_{c} \pm i\epsilon }  r^{\frac{d}{2}-m} \, S^{\mathrm{in}} = \lim_{r \rightarrow r_{c} \pm i\epsilon }  r^{\frac{d}{2}-m} \, S^{\mathrm{rev}} = S_{0} \, .
\end{split}
\end{equation}
Further, at these two boundaries, 
\begin{equation} 
\begin{split}
&\lim_{r \rightarrow r_{c} + i\epsilon } \sqrt{f}= 1\, , \quad \quad \qquad \lim_{r \rightarrow r_{c} - i\epsilon } \sqrt{f}= -1 \, .
\end{split}
\end{equation}
These boundary conditions immediately yield the following equations,
\begin{equation} 
\begin{split}
& \psi _{L}(\omega , \bm{k}) =  c(\omega , \bm{k}) +  h(\omega , \bm{k})
\\
& \psi _{R}(\omega , \bm{k}) =c(\omega , \bm{k}) -  h(\omega , \bm{k}) \,  e^{-\beta \omega } 
\end{split}
\end{equation}
These, in turn, are readily solved to give the full solution,
\begin{equation} 
\begin{split}
\Psi(\zeta , \omega , \bm{k}) = \, \, &S^{\mathrm{in}}(\zeta , \omega , \bm{k}) \, \Big[ n_{\omega }^{\mathrm{FD}} \, \psi _{L} + \big(1 - n^{\mathrm{FD}}_{\omega }\big) \, \psi _{R}\Big]
\\
& \ + \Big[\sqrt{f} \, \mathbb{\Gamma } + \frac{1}{\sqrt{f}} \, \mathbb{\Gamma }^{\dagger} \Big] \, S^{\mathrm{rev}}(\zeta , \omega , \bm{k}) \,  \Big[\big(1 - n_{\omega }^{\mathrm{FD}}\big) \, \big(\psi _{L} - \psi _{R}\big)\Big] \, e^{-\beta \omega \zeta }.
\end{split}
\end{equation}
Where $n^{\mathrm{FD}}_{\omega }$ is the familiar Fermi--Dirac factor \eqref{FD factor}. The solution can be alternately written as,
\begin{equation}
\Psi(\Ctor,\w, \bm{k})= -S^{\mathrm{in}}(\Ctor,\w,\bm{k}) \; \psiFb - \left[\sqrt{f} \, \mathbb{\Gamma } + \fr{1}{\sqrt{f}} \, \mathbb{\Gamma }^{\dagger}\right] \, S^{\mathrm{rev}} (\Ctor,\w, \bm{k}) \; \psiPb \; e^{\be \w (1-\Ctor)} \, ,
\end{equation}
where we have again introduced the retarded--advaced basis,
\begin{equation} \label{psi F P}
\psiFb = \FD \, (\psi_R-\psi_L)-\psi_R \ ,\qquad \psiPb = \FD \, (\psi_R-\psi_L) \, .
\end{equation}

We see that when the solution is written in terms of sources in the RA basis, the 
two combinations of sources precisely multiply the ingoing/quasi--normal bulk--to--boundary
propagator and the outgoing bulk--to--boundary propagator respectively. This fact can be exploited to argue why the generating function
of correlations computed from the bulk necessarily satisfy the SK collapse rules
as well as KMS conditions, i.e., the statement that the generating function cannot contain terms with only $\psiPb$ or terms with only $\psiFb$. 

The argument proceeds as follows:  if we think of the contributions to the generation function as arising from  Witten diagrams on grSK contour, the terms under question are either made entirely out of the ingoing propagators or made entirely out of the outgoing propagators. A diagram made completely out of ingoing propagators is analytic on the grSK contour and hence evaluates to zero, when the vertices are integrated all over the grSK contour, viz., the left and right contributions cancel out.  A diagram made completely out of outgoing propagators is also analytic on the grSK contour: the branch cut due to $\Ctor$ cancels
by frequency conservation, while the square root branch cut cancels because any non--zero
fermionic correlator has an even number of operator insertions. Hence, by a similar argument,
terms with only $\psiPb$ cannot occur in the generating function. This argument is a direct generalisation of a similar argument in the bosonic examples \cite{Chakrabarty:2019aeu,Jana:2020vyx}. 

Finally, in the Keldysh or average--difference basis \eqref{ad Sources} the  solution above takes the form,
\begin{equation} 
\begin{split}
\Psi (v, \zeta , \bm{k}) = \, \, &   S^{\mathrm{in}} \,\psi _{a} -\left\{ \left(n^{\mathrm{FD}}_{\omega }-\frac{1}{2}\right)  S^{\mathrm{in}} + n_{\omega }^{\mathrm{FD}} \, e^{\beta \omega (1-\zeta)}\Big[\sqrt{f} \, \mathbb{\Gamma } + \frac{1}{\sqrt{f}} \, \mathbb{\Gamma }^{\dagger} \Big] \, S^{\mathrm{rev}} \,  \right\} \psi _{d} \, .
\end{split}
\end{equation}

Having provided a derivation of the statistical factor for Dirac fermions, it is easy to see how this argument generalises to arbitrary spins.
The crucial fact in this regard is the statement that the outgoing Hawking modes suffer a Boltzmann suppression as well as a complex  boost of rapidity $2\pi i$ when they pass through the horizon cap 
region. The rapidity $2 \pi  i$ arises by comparing the Lorentz boost factor $\mathcal{T}$ \eqref{PT LLT} across the horizon cap --- the factor $2\pi i$ is essentially the jump in the rapidity $\vartheta = \ln f$ appearing in $\mathcal{T}$. It is then clear that for a field of spin $s$, the classical transmission amplitude for the outgoing modes across the horizon cap is given by,
\begin{equation}
   e^{-\beta \omega + 2 \pi i s} \ ,
\end{equation}

As we argued in the introduction , for integer $s$ this reduces to a simple Boltzmann weight. For half integer $s$, on the other hand, we get an additional phase factor of $-1$,
which then leads to a destructive interference between the ingoing and the outgoing modes, perfectly reproducing the Pauli exclusion
felt by half--integer fields. The classical physics of grSK saddles thus automatically reproduces the correct spin--statistics relation for the boundary correlators. 

\noindent

\section{Solution in gradient expansion}\label{sec:GradientExp}
In this section, we will describe how the Dirac equation can be solved explicitly in a derivative expansion.
We will begin by first discussing the massless case where the functions involved are simpler and then generalise
our discussion to the massive case.

\subsection{Massless gradient expansion}

In the massless case, the most general ingoing solution compatible with rotational invariance can be 
expanded in a derivative expansion as,
\begin{equation}\label{Massless Ansatz}
\begin{split}
\Psi = \fr{1}{r^{d/2}}\Big\{\mathds{1}+ C^{(1)}_{a} M^a \del_v &+ D^{(1)}_a M^{a}\Go{i}\del_i+ C^{(2)}_{a} M^a \del^2_v 
\\
&+ D^{(2)}_a M^{a}\del^2_i + F^{(2)}_a M^a \Go{i}\del_i \del_v+ \ldots \Big\}S_{0} \, \psi \ .
\end{split}
\end{equation}

Here  \(\psi = \psi(v,\bm{x}) \) is a (boundary) spinor that acs as the source at the boundary.  \(S_0 \) is the leading order ingoing propagator (up to factor of \(r^{-\fr{d}{2}} \)) and satisfies,

\begin{equation}\label{S0 Annihilation}
\mathbb{\Gamma } S_0 = 0  \ .
\end{equation}
In the above ansatz, \(\left\{C_a^{(n)},D_a^{(n)},F_a^{(n)},\ldots \right\} \) are functions of the radial coordinate \(\Ctor\) and \(M^a \in \{\mathds{1},\Go{\Ctor} \}\). Terms with $M^{a} = \Gamma ^{(v)}$ can eliminated using \eqref{S0 Annihilation}. 

The boundary conditions at infinity and the regularity condition at the horizon fix the ingoing solution.
\begin{equation}
\begin{split}
 \Psi_{\mathrm{in}} =&\  \fr{1}{r^{d/2}}\Bigg\{1-\fr{i\be}{2}\left(H\Go{\Ctor}- H_{(0)}\right)\Go{i}\;\del_i -\fr{\be^2}{2}\left[\widetilde{H}\Go{\Ctor}- \widetilde{H}_{(0)}\right]\Go{i}\;\del_i \del_v  \\
 &\qquad \qquad \qquad + \fr{\be^2}{8}\left(f H^2- f_c H^2_{(0)} \right)\del_i^2 + \fr{\be^2}{4} H_{(0)} \left(H\Go{\Ctor}- H_{(0)}\right)\del_i^2 \Bigg\} S_0 \psi \, .
\end{split}
\end{equation}
Here, \(H \) and \(\widetilde{H}\)  are functions defined by the differential equations and boundary conditions,
\begin{equation} \label{H Htilde}
\dCtor(H\sqrt{f})=\sqrt{f} \ ,\qquad \dCtor(\widetilde{H}\sqrt{f})=H\sqrt{f} \ ,\qquad H_{(0)} = H(\Ctor_c) \ ,\qquad \widetilde{H}_{(0)} = \widetilde{H}(\Ctor_c) \, .
\end{equation}
All these functions are regular in ingoing coordinates. The function \(H\) can be explicitly computed and is given by,
\begin{equation}
H(\Ctor) = \fr{1}{i \pi} \;{}_2F_1\left(\fr{1}{2},\fr{d-1}{d};\fr{3}{2};f(\Ctor)\right) \qquad \Rightarrow \qquad  H_{(0)} = -\fr{i}{ 2 \sqrt{\pi}} \fr{\Gamma(1/d)}{\Gamma(\fr{d+2}{2 \; d})}   \ .
\end{equation}

Given this ingoing solution in derivative expansion, the solution on the full grSK contour can be obtained from the expressions
derived in the last section. In the a-d basis the solution, up to second order in gradients, can be expressed as, 
\begin{equation}
\begin{split}
 \Psi = \ &   \fr{1}{r^{d/2}} \Bigg\{ \Bigg[ \mathds{1} + \fr{\be}{2}\left(H \Go{\Ctor}-H_{(0)} \right)\Go{i}k_i - \fr{\be^2}{2} \left(\widetilde{H}\Go{\Ctor}-\widetilde{H}_{(0)}\right)\Go{i}\;\w k_i 
 \\
 & \qquad \qquad \qquad \ - \fr{\be^2}{4}\bigg(H_{(0)} \left(H \Go{\Ctor}-H_{(0)}\right) +\fr{1}{2}\left(f H^2-H_{(0)}^2\right) \bigg)\bm{k}^2  \Bigg]S_0\psi_a  
 \\
 & \qquad  + \fr{1}{2}\Bigg[-\fr{\Go{\Ctor}}{\sqrt{f}} + \fr{\be}{2}\left(\mathds{1} + (2 \Ctor-1)\fr{\Go{\Ctor}}{\sqrt{f}} \right) \w 
  \\
  &\qquad \qquad \ \ \, + \fr{\be}{2}\left(H \sqrt{f} - H_{(0)} \fr{\Go{\Ctor}}{\sqrt{f}} \right)\Go{i}k_i  + \fr{\be^2}{2}\Ctor(1-\Ctor)\fr{\Go{\Ctor}}{\sqrt{f}}\;\w^2 
  \\
  &\qquad \qquad \ \ \, + \fr{\be^2}{8} \left( 2 H_{(0)} H \sqrt{f} + \left( f H^2 - 3 H_{(0)}^2\right)\fr{\Go{\Ctor}}{\sqrt{f}} \right) \bm{k}^2  
  \\
  &\qquad \qquad \ \ \,+ \fr{\be^2}{4}\Bigg(2 \left(\widetilde{H}\sqrt{f}-\widetilde{H}_{(0)}\fr{\Go{\Ctor}}{\sqrt{f}} \right) + (1-2\Ctor) \left(H\sqrt{f}-H_{(0)}\fr{\Go{\Ctor}}{\sqrt{f}}  \right) 
  \\
  &\qquad \qquad \qquad \qquad \qquad \qquad \qquad \qquad \quad+ \left( H\Go{\Ctor} -H_{(0)} \right) \Bigg)\Go{i}\;\w k_i\Bigg]S_0\psi_d\Bigg\}.
\end{split}
\end{equation}

 We can now calculate the on--shell action using the above solution --- as mentioned before, the contribution arises purely from boundary term of the Dirac action. When the dimension of the boundary theory is odd, $d = 2n +1$, using \eqref{odd d}, we have, 
\begin{equation}
\begin{split}
 r^{d} \; \bar{\Psi}\ProCtor_- \Psi\bigg|_{\Ctor=0}^{\Ctor=1} 
 &=\ -\overline{\psi_a}\left(\gamma^{(v)}+\be H_{(0)}\gamma^{(i)} k_i + \fr{\be^2}{2} H_{(0)}^2 \gamma^{(v)} \bm{k}^2 + \be^2 \widetilde{H}_{(0)} \gamma^{(i)} \w k_i\right)\psi_d \\
 &\ \quad + \overline{\psi_d}\left(\gamma^{(v)}-\be H_{(0)}\gamma^{(i)} k_i  + \fr{\be^2}{2} H_{(0)}^2 \gamma^{(v)} \bm{k}^2 + \be^2 \widetilde{H}_{(0)} \gamma^{(i)} \w k_i\right)\psi_a  \\
 &\ \quad + \fr{1 }{2} \overline{\psi}_d \left( \be \omega \gamma^{(v)}  \right)\psi_d  \ ,
 \\
 \\
 &=\ \overline{\psi_d}\gamma^{(v)}\psi_a-\overline{\psi_a}\gamma^{(v)}\psi_d  + \fr{\be \w }{2}  \;  \overline{\psi}_d \gamma^{(v)} \psi_d - \be H_{(0)} k_i\left( \overline{\psi_d} \gamma^{(i)}\psi_a +\overline{\psi_a}\gamma^{(i)} \psi_d  \right) \\
 &\ \quad -  \fr{\be^2}{2}H_{(0)}^2 \bm{k}^2    \left(\overline{\psi_a} \gamma^{(v)} \psi_d -\overline{\psi_d}\gamma^{(v)} \psi_a  \right)  - \be^2\widetilde{H}_{(0)}  \w k_i\left(\overline{\psi_a}\gamma^{(i)} \psi_d -\overline{\psi_d}\gamma^{(i)} \psi_a  \right).
\end{split}
\end{equation}

In the RL basis we have,
\begin{equation}
\begin{split}
 r^{d} \; \bar{\Psi}\ProCtor_- \Psi\bigg|_{\Ctor=0}^{\Ctor=1} =&\  \left(1-\fr{\be}{2}\w \right)\overline{\psi}_R \gamma^{(v)} \psi_L  - \left(1+\fr{\be}{2}\w \right)\overline{\psi}_L \gamma^{(v)} \psi_R  \\
 &\ + \fr{\be}{2}\; \overline{\psi}_R \left(\w \gamma^{(v)} - 2 H_{(0)} \gamma^{(i)} k_i \right)\psi_R + \fr{\be}{2} \; \overline{\psi}_L \left(\w\gamma^{(v)} +2 H_{(0)} \gamma^{(i)} k_i \right)\psi_L  \\
  &\ +\fr{\be^2}{2} \overline{\psi}_R \left( H_{(0)}^2 \gamma^{(v)} \bm{k}^2 + 2 \widetilde{H}_{(0)} \gamma^{(i)} \w k_i \right)\psi_L 
  \\
  & \qquad \qquad \qquad \qquad \qquad- \fr{\be^2}{2} \overline{\psi}_L \left( H_{(0)}^2 \gamma^{(v)} \bm{k}^2 + 2 \widetilde{H}_{(0)} \gamma^{(i)} \w k_i \right)\psi_R  \ .
\end{split}
 \end{equation} 
 
We quote the corresponding expression in even boundary dimensions $d=2n$ is (check Appendix \ref{Spinor and Gamma Conventions} for Clifford algebra conventions)

\begin{equation}
\begin{split}
    r^{d} \; \bar{\Psi}\ProCtor_- \Psi\bigg|_{\Ctor=0}^{\Ctor=1} &=\ \psi_{a}^{\dagger}\psi_d - \psi_{d}^{\dagger}\psi_{a} -\fr{\beta\w}{2}\psi_{d}^{\dag}\psi_{d}+ \beta H_{(0)}\left(\psi_{a}^{\dag} \Sigma^{(i)}k_i 
\psi_{d}+\psi_{d}^{\dag} \Sigma^{(i)}k_i 
\psi_{a} \right)  \\
&\quad +\fr{\beta^{2}}{2} \psi_{a}^{\dag}\left( H_{(0)}^{2} \bm{k}^2 + 2 \w k_{i} \widetilde{H}_{(0)} \Sigma^{(i)} \right)\psi_{d} -\fr{\beta^{2}}{2} \psi_{d}^{\dag}\left( H_{(0)}^{2} \bm{k}^2 + 2 \w k_{i} \widetilde{H}_{(0)} \Sigma^{(i)} \right)\psi_{a}  \ , \\
\\
&=\  \psi_{L}^{\dag} \left(1+\fr{\beta \w}{2} \right)\psi_{R}-\psi_{R}^{\dag} \left(1-\fr{\beta \w}{2} \right)\psi_{L} \\
&\quad +\fr{\beta}{2}\psi_{R}^{\dag}\left(2H_{(0)}\Sigma^{(i)}k_{i} -\w \right)\psi_{R} - \fr{\beta}{2}\psi_{L}^{\dag}\left(2H_{(0)}\Sigma^{(i)}k_{i} +\w \right)\psi_{L} \\
&\qquad +\fr{\beta^2}{2}\psi_{L}^{\dag} \left(H_{(0)} \bm{k}^{2} + 2 \widetilde{H}_{(0)} \w k_{i}\Sigma^{(i)} \right)\psi_{R} - \fr{\beta^2}{2}\psi_{R}^{\dag} \left(H_{(0)} \bm{k}^{2} + 2 \widetilde{H}_{(0)} \w k_{i}\Sigma^{(i)} \right)\psi_{L}
\end{split}
\end{equation}

\subsection{Massive gradient expansion}

In this section we will discuss the massive solution up to first order in derivative expansion. Note that while writing the following solution, we have  chosen the function  \( \sqrt{f}\) to assume positive values at the left boundary (the segment of contour with positive imaginary radius). In contrast to the massless solution, the structure of the ingoing solution in this case is more involved.
While a direct derivative expansion results in somewhat intricate differential equations even at first order in derivative expansion, we will show below that simpler equations can be obtained by a judicious parametrisation of the relevant functions.

We will begin by writing the zeroth order ingoing solution for the massive case in the form,
\begin{equation}
\begin{split}
 \Psi = \fr{1}{r^{\fr{d}{2}}}\fr{1}{\calNm}\Bigg\{ \mathcal{F}_{+} +\fr{\mathcal{F}_{-}}{\sqrt{f}} \Go{\Ctor} \Bigg\}S_0 \, \psi \, .
\end{split}
\end{equation}
The functions \(\mathcal{F}_{+} \) and \(\mathcal{F}_{-} \) are in turn defined via a function \(\mathcal{F}\) as,
\begin{equation}\label{F + - definition}
\mathcal{F}_{+} \equiv \fr{1}{2} \left( \mathcal{F} + \fr{1}{\mathcal{F}} \right) \ , \qquad \mathcal{F}_{-} \equiv \fr{1}{2} \left(\mathcal{F} - \fr{1}{\mathcal{F}} \right) \, ,
\end{equation}
where the function  \(\mathcal{F} \) satisfies, 
\begin{equation}
\dCtor \mathcal{F} = m \fr{i \be}{2} r \sqrt{f} \mathcal{F} \ ,\qquad  \mathcal{F}(\Ctor_h) =1 \, .
\end{equation}
The above conditions fix \(\mathcal{F} \) to be,
\begin{equation}\label{F definition}
\mathcal{F} = \left(\fr{1+\sqrt{f}}{1-\sqrt{f}}\right)^{m/d} \, .
\end{equation}
\(\calNm \) is a normalisation constant given by,
\begin{equation}
\calNm \equiv \lim_{r \to \infty} r^{-m} \mathcal{F} =    \fr{4^{\fr{m}{d}}}{ r_h^{m}} \, .
\end{equation}

Notice that across the horizon, \(\mathcal{F}\) and \(\mathcal{F}^{-1}\) are exchanged because the sign of the function \(\sqrt{f} \) reverses. This in turn reverses the sign of \(\mathcal{F}_{-}\) while the sign of \(\mathcal{F}_{+}\) remains invariant. Nevertheless, in the full zeroth order solution to the massive Dirac equation given above, no such square root branch cuts appear, thanks to the extra factor of \(\sqrt{f} \) multiplying \(\mathcal{F}_{-} \). 

At first order in derivatives we can write an ingoing solution of the form, 
\begin{equation}\label{Massive ingoing solution first order}
\begin{split}
	\Psi =&\  \fr{1}{r^{\fr{d}{2}}}\fr{1}{\calNm}\Bigg\{ \left( \mathcal{F}_{+} + \mathcal{F}_{-} \, \frac{\Go{\Ctor}}{\sqrt{f}}\right)\left(1+ \fr{i \be}{2} \mathcal{W}_{+}\;\del_v +  \fr{i \be}{2} \mathcal{K}_{+}\;\Go{i}\del_i\right) 
	\\
	&\qquad \qquad \qquad \qquad - \left( \mathcal{F}_{-} + \mathcal{F}_{+} \, \frac{\Go{\Ctor}}{\sqrt{f}} \right) \left(\fr{i \be}{2} \mathcal{W}_{-}\;\del_v +  \fr{i \be}{2} \mathcal{K}_{-}\;\Go{i}\del_i\right) \Bigg\}S_0 \psi \, .
\end{split}
\end{equation}
Here the functions \(\mathcal{W}_{\pm} \) and  \(\mathcal{K}_{\pm} \) are solutions to the differential equations,
\begin{equation}
\begin{split} \label{W K definitions}
	\dCtor \mathcal{W}_{+} =&\  2 \; (\mathcal{F}_{-})^2 \, ,\qquad \quad \;   \dCtor \mathcal{W}_{-} = 2 \; \mathcal{F}_{+} \mathcal{F}_{-} \, ,  \\ 
	\dCtor \mathcal{K}_{+} =&\  2 \sqrt{f} \mathcal{F}_{+} \mathcal{F}_{-} \, ,\qquad \dCtor \mathcal{K}_{-} = \sqrt{f} \; \left((\mathcal{F}_{+})^2 +(\mathcal{F}_{-})^2\right)   \, .\\ 
\end{split}
\end{equation}
They are uniquely fixed by the boundary conditions
\begin{equation}
\begin{split}
	\lim_{\Ctor \to 0}( \mathcal{W}_{+} - \mathcal{W}_{-}) &=0 \, , \qquad \qquad \lim_{\Ctor \to 0}( \mathcal{K}_{+} - \mathcal{K}_{-})  =0 \, \\
	\lim_{\Ctor \to \Ctor_{h}}\mathcal{W}_{-} &=0 \, , \qquad \qquad \qquad \quad \, \lim_{\Ctor \to \Ctor_{h}}\mathcal{K}_{-} =0 \, .
\end{split}
\end{equation}

In the above, exactly like the functions at zeroth order, the functions with subscript `$+$' remain the same as we traverse the horizon cap. The functions with subscript `$-$', on the other hand, pick up a minus sign because of a square root branch cut. We can write an integral representation for \(\mathcal{W}_{\pm} \) and  \(\mathcal{K}_{\pm} \) as,
\begin{equation}
\begin{split}
\mathcal{W}_{+}(\Ctor) =&\ 2 \int_{0}^{\zeta}(\mathcal{F}_{-})^2 \;d\hat{\Ctor} - 2 \int^{\Ctor_h}_{0}\mathcal{F}_{+} \mathcal{F}_{-} \;d\hat{\Ctor} \, , 
\\
\mathcal{W}_{-}(\Ctor) =&  2 \int_{\Ctor_h}^{\Ctor}\mathcal{F}_{+} \mathcal{F}_{-} \;d\hat{\Ctor} \, ,
\\ 
\mathcal{K}_{+}(\Ctor) =&\ 2 \int_{0}^{\zeta} \sqrt{f} \mathcal{F}_{+} \mathcal{F}_{-} \;d\hat{\Ctor} - 2 \int^{\Ctor_h}_{0}\sqrt{f} \; \left((\mathcal{F}_{+})^2 +(\mathcal{F}_{-})^2\right) \;d\hat{\Ctor} \, , 
\\ 
\mathcal{K}_{-}(\Ctor) =&\  \int_{\Ctor_h}^{\Ctor} \sqrt{f} \; \left((\mathcal{F}_{+})^2 +(\mathcal{F}_{-})^2\right) \;d\hat{\Ctor} \,  .\\ 
\end{split}
\end{equation}
The reader can check that these functions satisfy the differential equations and the horizon jump conditions mentioned above.

The $\mathcal{W}_{-}$ and $\mathcal{K}_{-}$ functions vanish at the horizon, as can be checked by computing the following limits. 
\begin{equation}\label{Horizon values}
\begin{split}
	\lim_{\Ctor\to\Ctor_h} \frac{1}{\sqrt{f}} \, \mathcal{W}_{-}(\Ctor)=&\  \lim_{\Ctor\to\Ctor_h} \fr{1}{\sqrt{f}}\int_{\Ctor_h}^{\Ctor} 2 \; \mathcal{F}_{+} \mathcal{F}_{-} d\hat{\Ctor} = \fr{1}{i \pi} \fr{4}{d} \,  m \, , \\
	\lim_{\Ctor\to\Ctor_h} \frac{1}{\sqrt{f}} \, \mathcal{K}_{-}(\Ctor)=&\  \lim_{\Ctor\to\Ctor_h} \fr{1}{\sqrt{f}}\int_{\Ctor_h}^{\Ctor} \sqrt{f} \; \left((\mathcal{F}_{+})^2 +(\mathcal{F}_{-})^2\right) d\hat{\Ctor}  =  \fr{1}{i \pi}  \,  .
\end{split}
\end{equation}

As in the massless case, the solution on the grSK saddle can be obtained by using the ingoing solution. In the a-d basis, we get (upto first order in derivatives), 
\begin{equation}
\begin{split}
  \Psi =&\  \fr{1}{r^{\fr{d}{2}}}\fr{1}{\calNm}\Bigg\{ \Bigg( \left[ \mathcal{F}_{+} +\fr{\mathcal{F}_{-}}{\sqrt{f}} \Go{\Ctor} \right] \left(1+ \fr{\be \w}{2}\mathcal{W}_{+}-\fr{\be k_i}{2} \mathcal{K}_{+}\Go{i}\right) \\
  &\ \qquad \qquad  -\left[ \mathcal{F}_{-} + \fr{ \mathcal{F}_{+}}{\sqrt{f}} \Go{\Ctor} \right] \left( \fr{\be\w}{2}\mathcal{W}_{-}- \fr{\be k_i}{2}\mathcal{K}_{-}\Go{i} \right)\Bigg)S_0\psi_a  \\
  &\ \qquad \qquad -\fr{1}{2}\Bigg(\left[\mathcal{F}_{-} + \fr{\mathcal{F}_{+}}{\sqrt{f}}\Go{\Ctor}\right]\left(1+\fr{\be \w}{2}\left(1 -2\Ctor - \mathcal{W}_{+} \right)+\fr{\be k_i}{2}\mathcal{K}_{+}\Go{i} \right) \\ 
  & \qquad \qquad \qquad \quad \ \ + \left[\mathcal{F}_{+}+\fr{\mathcal{F}_{-}}{\sqrt{f}}\Go{\Ctor} \right]\left(\fr{\be \w}{2} \left( \mathcal{W}_{-}-1\right)- \fr{\be k_i}{2} \mathcal{K}_{-} \Go{i}  \right)   \Bigg)S_0\psi_d\Bigg\} .  
\end{split}
\end{equation}

In order to calculate the influence phase, we need to evaluate on--shell action with appropriate counterterms. To do this, we need the boundary limit of the above solution. We start by defining the following boundary limits of the functions,
\begin{equation}
\begin{split}
\lim_{r \to r_c + i \epsilon}\mathcal{F}=&\  \mathcal{F}_{c} \, ,\qquad \lim_{r \to r_c + i \epsilon}\mathcal{W}_{\pm}=\mathcal{W}_{c} \, ,\qquad  \quad \lim_{r \to r_c + i \epsilon}\mathcal{K}_{\pm}=\mathcal{K}_{c} \, , \\
\lim_{r \to r_c - i \epsilon}\mathcal{F}=&\  \fr{1}{\mathcal{F}_{c}}\, ,\qquad \lim_{r \to r_c - i \epsilon}\mathcal{W}_{\pm}= \pm \mathcal{W}_{c} \, ,\qquad \lim_{r \to r_c - i \epsilon}\mathcal{K}_{\pm}=\pm \mathcal{K}_{c} \, , \\
\end{split}
\end{equation}
More explicitly, we have,
\begin{equation}\label{F c definition}
 \mathcal{F}_{c} =  \left( 4\fr{r_c^d}{r_h^d}  \right)^{|m|/d} \implies \calNm r_c^{m} =  \mathcal{F}_{c} \, ,
\end{equation}

In terms of this data, the boundary influence phase can be computed by evaluating the on--shell Dirac action using our solution. In the a-d basis, when the dimension of the boundary theory is odd $d=2n+1$, we obtain,
\begin{equation}
\begin{split}
 \bigg[ r^{d} \bar{\Psi}\ProCtor_{-}\Psi\bigg]_{\Ctor=0}^{\Ctor=1} 
 =&\ \left(\fr{r_h^{2m}}{16^{(m/d)}}\right)\Big\{-\bar{\psi}_a\left((1-\be \w\;\mathcal{W}_{c})\gamma^{(v)} + \be k_i  \; \mathcal{K}_{c} \gamma^{(i)}\right)\psi_d  \\
 & \qquad \qquad \qquad \ \, + \bar{\psi}_d \left( \left(1+\be \w\mathcal{W}_{c} \right)\gamma^{(v)} - \be k_i \mathcal{K}_{c} \gamma^{(i)} \right)\psi_a + \fr{\be \w}{2}\bar{\psi}_d \gamma^{(v)} \psi_d \Big\}\, .
\end{split}
\end{equation}
Similarly, when $d = 2n$, we have,
\begin{equation}\label{key}
\begin{split}
 \bigg[ r^{d} \bar{\Psi}\ProCtor_{-}\Psi\bigg]_{\Ctor=0}^{\Ctor=1} 
=&\ \left(\fr{r_h^{2m}}{16^{(m/d)}}\right)\Big\{\psi^{\dagger}_a\left( (1-\be \w\;\mathcal{W}_{c}) - \be k_i  \, \mathcal{K}_{c} \, \Sigma ^{(i)}\right)\psi_d  \\
& \qquad \qquad \qquad \ \, - \psi^{\dagger}_d \left( \left(1 + \be \w \, \mathcal{W}_{c} \right) + \be k_i \,  \mathcal{K}_{c} \,  \Sigma ^{(i)} \right)\psi_a - \fr{\be \w}{2} \, \psi^{\dagger}_d  \psi_d \Big\}\, .
\end{split}
\end{equation}
The relevant finite coefficients are given by series expansions involving the incomplete Beta functions (or digamma functions), as described in Appendix \ref{app:Asymp}.
\begin{equation}
\begin{split}
\Big\{\fr{4 \pi i }{ 16^{\fr{m}{d}} d} \mathcal{W}_{c} \Big\}_{r^{0}} =&\ - \fr{1 }{ 16^{\fr{m}{d}}} \Bigg[\fr{ 4^{\fr{1}{d}+1} m}{4m^2-1} +\fr{2}{d} \left\{\psi\left(\fr{d+1-2m}{d}\right)-\psi\left(\fr{d+1+2m}{d}\right)\right\} \\
&\quad \qquad  - \sum_{j=1}^{\infty}  \left[  B\left(j,\fr{2}{d},-1 \right) - 2^{\fr{2}{d}-1}\fr{(-)^{j}}{j} \right] \fr{1}{B\left(j,\fr{2}{d} ,1\right)} \fr{8m}{4m^2-(1+dj)^2} \Bigg]  \\
\Big\{\fr{4 \pi i }{ 16^{\fr{m}{d}} d} \mathcal{K}_{c}\Big\}_{r^{0}} =&\  4^{\fr{1-2m}{d}}   \fr{\Gamma\left(\fr{1+2m}{d} 
\right) \Gamma\left(\fr{1-2m}{d} 
\right)}{d \; \Gamma\left(\fr{2}{d} 
\right)} \ .
\end{split}
\end{equation}
This gives an explicit expression for the influence phase for any value of the mass of the bulk Dirac fermion.

\section{Discussion and future directions}\label{sec:DiscConc}
In this paper, we derive the open effective field theory of a fermion probing a thermal CFT bath using holography. The probe sources a (fermionic) single trace primary. The influence phase of this probe is computed by solving the bulk Dirac equation on the gravitational Schwinger--Keldysh (grSK) geometry, and evaluating the bulk on--shell action on this solution. This procedure correctly reproduces the Fermi--Dirac factors expected from the CFT side.

It is difficult to find exact analytic solutions of the Dirac equation in the black brane background for $d>2$. However, we show, for the first time, that analytic expressions can be obtained for the ingoing solution in a boundary derivative expansion. Explicit functions are presented for the massless case up to second order in derivates. For massive fields, we stop at first derivative order. These results are checked against the exact expressions available in $d = 2$. Evaluating the bulk on--shell action, we obtain an explicit expression for the influence phase in derivative expansion.

An immediate generalisation of our work is to turn on a  finite chemical potential in the CFT bath. In the bulk, this corresponds to probing charged black brane backgrounds. We consider this scenario in an accompanying paper \cite{Loganayagam:2020iol}.
 
 Another interesting direction is to analyse the corrections to the probe effective action due to interactions in the bulk. For instance, one could introduce Yukawa couplings between the bulk fermion and another bulk scalar field. Other possibilities include four fermion coupling, as well as minimal and non--minimal couplings to a gauge field. These cases all capture different physical effects that appear in interacting fermionic open EFTs. An interesting question in this context is to understand renormalisation in such EFTs. Previous studies of interacting scalars in the grSK background already reveal the interesting phenomenon of `source' renormalisation \cite{Jana:2020vyx}, and it would be nice to study the fermionic analogue.

\section*{Acknowledgements}
It is with great pleasure that we thank Bidisha Chakraborty, Joydeep Chakravarty, Greg Henderson, Chris Herzog, Veronika Hubeny, Chandan Jana, Aswin P. M., Suvrat Raju, Mukund Rangamani, Shivam Sharma, Aninda Sinha and Spenta Wadia for useful discussions. RL would like to thank the organizers of the UC Davis String Theory Seminars where a part of this work was presented. KR is supported by the Rhodes Trust via a Rhodes Scholarship. The authors would like to acknowledge their debt to the people of India for their sustained and generous support to research in the basic sciences.
 \appendix

\section{Spinors and Clifford algebras: conventions and representations}\label{Spinor and Gamma Conventions}
In this appendix, we delineate our conventions for the spinors and Clifford algebras used in the main text. We organize our discussion from the perspective of the boundary conformal field theory, first describing the conventions for the boundary spinors and boundary Gamma / Sigma matrices, and then explaining how the corresponding objects in the bulk are defined. The exposition presented here and the choice of basis that we use differ slightly from the standard references on the subject \cite{Iqbal:2009fd,Polchinski:1998rr,West:2012vka,Freedman:2012zz,Ortin:2015hya}, but it is one that we find useful for our purposes. Of course, none of the results quoted in the text depend on a particular representation of these matrices.

\subsection{Spinor and Clifford algebra conventions} 
The Gamma matrices that we use satisfy the Clifford algebra,
\begin{equation} 
\begin{split}
\Big\{\Gamma ^{a}, \, \Gamma ^{b}\Big\} = 2 \, \eta ^{ab} \, \mathds{1} ,
\end{split}
\end{equation}
where $\eta _{ab} = \mathrm{diag} \, (-1, \, +1, \ldots, \, +1)$. In particular, the lone time--like Gamma matrix $\Gamma ^{(v)}$ squares to $-\mathds{1}$, while the space--like Gamma matrices are idempotent. We will always choose $\Gamma ^{(v)}$ to be \emph{anti}--Hermitian and all the rest to be Hermitian. Recall that in a $(d+1)$--dimensional spacetime, these are $2^{k} \times 2^{k}$ dimensional matrices, with $k$ given by,
\begin{equation}\label{Spinor dimension}
\begin{split}
k = \Big\lfloor \frac{d+1}{2}\Big\rfloor  . 
\end{split}
\end{equation}

Thus, the Dirac spinors in this spacetime also have $2^{k}$ components. As we had noted before, the bulk and boundary spinors can have a different number of components, depending on the dimension of the spacetime we are in. In this regard, we adopt the following conventions: the spinors in the bulk will always be represented as Dirac spinors. When the dimension of the boundary is odd, i.e., $d = 2n +1$, the spinors of the CFT are Dirac spinors of dimension $2^{n}$. When the dimension of the boundary spacetime is even, i.e., $d = 2n$, we will use Weyl spinors for the CFT, viz., chiral or anti--chiral  spinors of dimension $2^{n-1}$. With these choices, in every dimension the bulk spinors always have twice as many components as the boundary spinors do.

We choose a basis for our Gamma matrices that plays well with the above conventions. When $d = 2n$, the boundary spinors are Weyl spinors that transform under the covariant Pauli (or Sigma) matrices $(\Sigma , \overline{\Sigma })$ given by
\begin{equation}\label{Sigma matrices}
\begin{split}
\Sigma ^{(\mu )} = \left(\mathds{1}, \sigma ^{(i)}\right) \, ; \qquad \overbar{\Sigma }^{(\mu )} = \left(\mathds{1}, - \sigma ^{(i)}\right) \ .
\end{split}
\end{equation}
In turn, $\sigma ^{i}$ are the generalized Pauli matrices that satisfy,
\begin{equation}\label{Pauli algebra} 
\begin{split}
\big\{\sigma ^{(i)},\sigma ^{(j)}\big\} = 2 \, \delta ^{(i)(j)} \, \mathds{1}  .
\end{split}
\end{equation}
Correspondingly, we choose the the Gamma matrices in the bulk to be,
\begin{equation} 
\begin{split}
\Gamma ^{(\zeta )} = \begin{pmatrix}
\mathds{1} & 0 \\
0 & - \mathds{1}
\end{pmatrix}\, , \qquad \Gamma ^{(\mu )} = \begin{pmatrix}
0 & \Sigma ^{(\mu )} \\
- \overbar{\Sigma }^{(\mu )} & 0
\end{pmatrix}  .
\end{split}
\end{equation}
It is easily checked that these Gamma matrices satisfy the Clifford algebra. Furthermore, because the Sigma matrices are all Hermitian, $\Gamma ^{(v)}$ is an anti--Hermitian matrix, while all the rest are Hermitian. Finally, the `lightcone' combinations of Gamma matrices $\mathbb{\Gamma }$ and $\mathbb{\Gamma }^{\dagger}$ \eqref{BBGamma} and the constant bulk to boundary matrix $S_{0}$ \eqref{S0} that we introduced in the main text, read, 
\begin{equation} 
\begin{split}
&\mathbb{\Gamma } = \frac{1}{2} \, \Big(\Gamma ^{(v)} + \Gamma ^{(\zeta )}\Big) = \frac{1}{2} \begin{pmatrix}
 \mathds{1} & \Sigma ^{(v)} \\
- \overbar{\Sigma }^{(v)} & -\mathds{1}
\end{pmatrix} = \frac{1}{2} \begin{pmatrix}
\mathds{1} & \mathds{1} \\
- \mathds{1} & -\mathds{1}
\end{pmatrix}  , \qquad S_{0} = \begin{pmatrix}
\mathds{1} \\
- \mathds{1}
\end{pmatrix}  .
\end{split}
\end{equation}  

When $d = 2n-1$, the boundary spinors are Dirac spinors that transform under the boundary Gamma matrices, which we will denote by $\gamma ^{(\mu )}$. In this case, the Gamma matrices for the bulk spinor can be chosen to be,
\begin{equation} 
\begin{split}
 \Gamma ^{(\zeta )} = \begin{pmatrix}
\mathds{1} & 0
\\
0 & -\mathds{1}
\end{pmatrix} \, , \qquad \Gamma ^{(\mu )} = \begin{pmatrix}
0 & \gamma ^{(\mu )} \\
\gamma ^{(\mu )} & 0
\end{pmatrix}  .
\end{split}
\end{equation}
Once again, it is easy to see that the above Gamma matrices satisfy the Clifford algebra, and that they inherit the correct Hermiticity properties from the boundary $\gamma $'s, i.e., $\Gamma ^{(v)}$ is an anti--Hermitian matrix, and the rest of the $\Gamma ^{(a)}$ are Hermitian. In these dimensions, the matrices $\mathbb{\Gamma }$, $\mathbb{\Gamma }^{\dagger}$ and $S_{0}$ then take the form, 
\begin{equation} 
\begin{split}
&\mathbb{\Gamma } = \frac{1}{2} \begin{pmatrix}
\mathds{1} & \gamma ^{(v)} \\
\gamma ^{(v)} & -\mathds{1}
\end{pmatrix}  , \qquad S_{0} = \begin{pmatrix}
\mathds{1} \\ 
\gamma ^{(v)}
\end{pmatrix}  .
\end{split}
\end{equation}

\noindent
\subsection{Pauli matrices and explicit representations} \label{Generalized Pauli Matrices.}
Having described how to construct the bulk Gamma matrices from the $(\Sigma , \, \overbar{\Sigma })$ and $\gamma $ matrices, we now turn to providing explicit representations for these boundary matrices. We will do this using the generalized Pauli matrices. 

As the name suggest, the generalized Pauli matrices are set of Hermitian matrices that  generalize the standard $2 \times 2$ Pauli matrices $(\tau _{1}, \, \tau _{2}, \, \tau _{3})$ to all odd dimensions. As we will soon see, we can always choose them to be real and symmetric, or imaginary and anti--symmetric. In $p = 2n-1$ dimensions, they are $2^{n-1} \times 2^{n-1}$ matrices that satisfy the Euclidean Clifford algebra,  
\begin{equation} 
\begin{split}
\big\{\sigma_{p} ^{(i)}, \sigma_{p} ^{(i)}\big\} = 2\, \delta ^{(i)(j)} \, \mathds{1}.
\end{split}
\end{equation}
We organize these matrices into four different classes depending on their dimension: $p = 8k+1$, $p= 8k+3$, $p = 8k+5$, $p = 8k+7$, and describe each of their explicit constructions separately. In each case, we will build them out of the standard Pauli matrices $(\tau _{1}, \, \tau _{2}, \, \tau _{3})$.

\noindent
\underline{$p = 8k+1$} \\
The very simplest case is of course $p =1$, where the Pauli `matrix' $\sigma _{1}^{(1)}$ is just the number 1. The next member of this family is $p =9$, where \emph{all} the Pauli matrices can be chosen to real, symmetric matrices as follows,
\begin{equation} 
\begin{split}
&\sigma _{9}^{(1)} = \tau _{1} \otimes \mathds{1} \otimes \mathds{1} \otimes \mathds{1}  \, ,
\\
&\sigma _{9}^{(2)} = \varepsilon \otimes \mathds{1} \otimes \tau _{3} \otimes \varepsilon \, , \quad  \sigma _{9}^{(3)} =\varepsilon \otimes \mathds{1} \otimes \tau _{1} \otimes \varepsilon,
\\
&\sigma _{9}^{(4)} = \varepsilon  \otimes \tau _{3} \otimes \varepsilon \otimes \mathds{1} \, , \quad \sigma _{9}^{(5)} = \varepsilon \otimes \tau _{1} \otimes \varepsilon \otimes \mathds{1},
\\
&\sigma _{9}^{(6)} = \varepsilon \otimes \varepsilon \otimes \mathds{1} \otimes \tau _{3} \, , \quad \sigma _{9}^{(7)} = \varepsilon \otimes \varepsilon \otimes \mathds{1} \otimes \tau _{1}, \quad 
\\
&\sigma _{9}^{(8)} = \epsilon  \otimes \epsilon  \otimes \epsilon \otimes \epsilon \, , \quad \ \ \, \sigma _{9}^{(9)} =  \tau _{3} \otimes \mathds{1} \otimes \mathds{1} \otimes \mathds{1}.
\end{split}
\end{equation}
We have adopted the shorthand $\varepsilon = i \tau _{2}$ to make the reality of these matrices manifest. With the above two cases defined, we can now recursively define the rest of the members of this family.
\begin{equation} 
\begin{split}
&\sigma _{8k+9}^{(i)} = \sigma _{9}^{(i)}  \otimes ^{4k} \mathds{1} \, ,  \ \qquad  i \in \{1, \ldots, 8\} ;
\\
&\sigma _{8k+9}^{(8+j)} = \sigma ^{(9)}_{9} \otimes \sigma _{8k+1}^{(j)} \, , \quad j \in \{1, \ldots , 8k+1\} .
\end{split}
\end{equation}

\noindent
\underline{$p = 8k+3$} \\
The matrices in these dimensions are defined as follows.
\begin{equation} 
\begin{split}
&\sigma _{8k+3}^{(1)} = \tau _{1}  \otimes ^{4k} \mathds{1} \, ,
\\
&\sigma _{8k+3}^{(2)} = \tau _{2}  \otimes ^{4k} \mathds{1} \, ,
\\
&\sigma _{8k+3}^{(2+j)} = \tau _{3} \otimes \sigma _{8k+1}^{(j)} \, , \quad j \in \{1, \ldots , 8k+1\} .
\end{split}
\end{equation}
Unsurprisingly, when $p=3$, these are just the standard Pauli matrices.

\noindent
\underline{$p = 8k+5$} \\
These matrices can be written down as follows.
\begin{equation} 
\begin{split}
&\sigma _{8k+5}^{(1)} = i \, \tau _{1}  \otimes \varepsilon \otimes ^{4k} \mathds{1} \, ,
\\
&\sigma _{8k+5}^{(2)} = i \, \tau _{3} \otimes \varepsilon \otimes ^{4k} \mathds{1} \, ,
\\
&\sigma _{8k+5}^{(3)} = \varepsilon \otimes \varepsilon \otimes ^{4k} \mathds{1} \, ,
\\
&\sigma _{8k+5}^{(4)} = \mathds{1} \otimes \tau _{3} \otimes ^{4k} \mathds{1} \, ,
\\
&\sigma _{8k+5}^{(4+j)} = \mathds{1} \otimes \tau _{1} \otimes \sigma _{8k+1}^{(j)} \, , \quad j \in \{1, \ldots , 8k+1\} .
\end{split}
\end{equation}

\noindent
\underline{$p = 8k+7$} \\
All the matrices in these dimensions can be chosen to be purely imaginary and anti--symmetric.
\begin{equation} 
\begin{split}
&\sigma _{8k+7}^{(1)} = i \, \mathds{1} \otimes  \tau _{3}  \otimes \varepsilon \otimes ^{4k} \mathds{1} \, ,
\\
&\sigma _{8k+7}^{(2)} = i \, \mathds{1} \otimes  \tau _{1} \otimes \varepsilon \otimes ^{4k} \mathds{1} \, ,
\\
&\sigma _{8k+7}^{(3)} = i \, \tau _{3}  \otimes \varepsilon \otimes \mathds{1} \otimes ^{4k} \mathds{1} \, ,
\\
&\sigma _{8k+7}^{(4)} = i \, \tau _{1}  \otimes \varepsilon \otimes \mathds{1} \otimes ^{4k} \mathds{1} \, ,
\\
&\sigma _{8k+7}^{(5)} = i \, \varepsilon \otimes \mathds{1} \otimes  \tau _{3} \otimes ^{4k} \mathds{1} \, ,
\\
&\sigma _{8k+7}^{(6)} =  i \, \varepsilon \otimes \mathds{1} \otimes  \tau _{1} \otimes ^{4k} \mathds{1} \, ,
\\
&\sigma _{8k+7}^{(6+j)} = i \, \varepsilon \otimes  \varepsilon \otimes \varepsilon \otimes \sigma _{8k+1}^{(j)} \, , \quad j \in \{1, \ldots , 8k+1\} .
\end{split}
\end{equation}

With these matrices explicitly laid out, we can now construct the boundary matrices in the AdS/CFT correspondence. For an even--dimensional boundary, $d = 2n = p+1$, the Sigma matrices are constructed as,
\begin{equation} 
\begin{split}
&\Sigma ^{(v)}_{d} = \overbar{\Sigma }_{d}^{(v)} = \mathds{1} \, , \qquad \quad \Sigma ^{(i)}_{d} = -\overbar{\Sigma }^{(i)}_{d} = \sigma ^{(i)}_{d-1} \, , \ \  i \in \{1, \ldots, \, d-1\}
\end{split}
\end{equation}
For an odd--dimensional boundary, $d = 2n+1=p+2$, the boundary Gamma matrices can be chosen as follows,
\begin{equation} 
\begin{split}
&\gamma _{d}^{(v)} = \begin{pmatrix}
0 & \mathds{1} \\ 
-\mathds{1} & 0
\end{pmatrix} \, ,
\qquad \quad \gamma _{d}^{(\mu )} = \begin{pmatrix}
0 & \sigma _{d-2}^{(\mu )} \\ 
\sigma _{d-2}^{(\mu )} & 0
\end{pmatrix} \, , 
\qquad \quad \gamma ^{(d-1)}_{d} = \begin{pmatrix}
\mathds{1} & 0 \\
0 & -\mathds{1}
\end{pmatrix} \, , 
\end{split}
\end{equation}
where, $\mu  \in \{v, \, 1, \ldots, \, d-2\}$. The corresponding matrices in the bulk can now be immediately constructed from our previous discussion. We end this appendix by listing out explicit representations of these matrices in a few low dimensional cases.

\subsection*{\underline{$d=2$}}
\begin{equation}\label{Matrices in d=2}
\begin{split}
&\Sigma _{2}^{(v)} = 1 \, , \quad \Sigma _{2}^{(1)} = 1 \, ; \quad \overbar{\Sigma }_{2}^{(v)} = 1 \, , \quad \overbar{\Sigma }_{2}^{(1)} = -1 \, ; \quad S_{0} = \begin{pmatrix}
1 \\ 
-1
\end{pmatrix} \, ;
\\
&\Gamma _{3}^{(v)} = i  \tau _{2} \, , \quad \Gamma _{3}^{(1)} = \tau _{1} \, , \quad \Gamma _{3}^{(\zeta )} = \tau _{3} \, .
\end{split}
\end{equation}

\subsection*{\underline{$d=3$} }
\begin{equation} 
\begin{split}
&\gamma _{3}^{(v)} = i\tau _{2} \, , \quad \gamma _{3}^{(1)} = \tau _{1} \, , \quad \gamma _{3}^{(2)} = \tau _{3} \, ; \quad S_{0} = \begin{pmatrix}
\mathds{1} \\
i\tau _{2}
\end{pmatrix} \, ;
\\
&\Gamma _{4}^{(v)} = \begin{pmatrix}
0 & i \tau _{2} \\
i \tau _{2} & 0
\end{pmatrix} \, , \quad \Gamma _{4}^{(1)} = \begin{pmatrix}
0 & \tau _{1} \\
\tau _{1} & 0
\end{pmatrix} , \quad \Gamma _{4}^{(2)} = \begin{pmatrix}
0 & \tau _{3} \\
\tau _{3} & 0
\end{pmatrix}  , \quad \Gamma _{4}^{(\zeta )} = \begin{pmatrix}
\mathds{1} & 0 \\
0 & -\mathds{1}
\end{pmatrix} \, .
\end{split}
\end{equation}

\subsection*{\underline{$d=4$} }
\begin{equation} 
\begin{split}
&\Sigma _{4}^{(v)} = \mathds{1} \, , \quad \Sigma _{4}^{(1)} = \tau _{1} \, , \quad \Sigma _{4}^{(2)} = \tau _{2} \, , \quad \Sigma _{4}^{(\zeta )} = \tau _{3} \, ;
\\
&\overbar{\Sigma }_{4}^{(v)} = \mathds{1} \, , \quad \overbar{\Sigma }_{4}^{(1)} = -\tau _{1} \, , \quad \overbar{\Sigma }_{4}^{(2)} = -\tau _{2} \, , \quad \overbar{\Sigma }_{4}^{(\zeta )} = -\tau _{3} \, ; \quad S_{0} = \begin{pmatrix}
\mathds{1} \\
-\mathds{1} \\
\end{pmatrix} \, ;
\\
&\Gamma _{5}^{(v)} = \begin{pmatrix}
0 & \mathds{1} \\
-\mathds{1} & 0
\end{pmatrix} \, , \quad \Gamma _{5}^{(1)} = \begin{pmatrix}
0 & \tau _{1} \\
\tau _{1} & 0
\end{pmatrix} \, , \quad \Gamma _{5}^{(2)} = \begin{pmatrix}
0 & \tau _{2} \\
\tau _{2} & 0
\end{pmatrix} \, , \quad \Gamma _{5}^{(3)} = \begin{pmatrix}
0 & \tau _{3} \\
\tau _{3} & 0
\end{pmatrix} \, ,
\\
&\Gamma _{5}^{(\zeta )} = \begin{pmatrix}
\mathds{1} & 0 \\
0 & -\mathds{1}
\end{pmatrix} \, .
\end{split}
\end{equation}

\section{The analysis in $d = 2$} \label{app:Exact Solution}
The situation in $d=2$ is simple enough that it admits an exact solution. As will see,  it is possible to obtain a closed form boundary--to--bulk infalling Green's function, and consequently an influence functional that is valid to all orders in gradient expansion. While an important and illuminative calculation in its own right, it is also a useful laboratory to check the validity of the perturbative study that we carried out in $\S$\ref{sec:GradientExp}.

This appendix is structured as follows: we begin by solving for the infalling boundary--to--bulk Green's function and describing the construction of the full boundary--to--bulk Green's function. We then move on to computing the influence functional, and commenting on the retarded and advanced Green's functions and the fluctuation--dissipation theorem of the boundary 2$d$ CFT that we obtain. Finally, we expand the infalling Green's function and show that they're consistent with the results of the gradient expansion.

\subsection{The boundary--to--bulk Green's function}
Following the main text, we will first solve for the infalling propagator in AdS$_{3}$ and then lift the solution on to the grSK saddle. The Dirac wave equation in the BTZ spacetime (of which AdS$_{3}$ is a special case) was first worked out in \cite{Das:1999pt} --- the solution presented here is adapted to our choice of tangent frame, and is motivated by structure of the functions in the gradient expansion ($\S$\ref{sec:GradientExp}). 

The metric for AdS$_{3}$ written in standard Eddington--Finkelstein coordinates is,
\begin{equation} 
\begin{split}
\mathrm{d}s^{2} &= -r^{2} \, \bigg( 1-\frac{r_{h}^{2}}{r^{2}}\bigg) \, \mathrm{d}v^{2} + 2 \, \mathrm{d}v \, \mathrm{d}r +r^{2} \, \mathrm{d}x^{2} .
\end{split}
\end{equation}
We continue to terminate this space--time at a finite radius $r = r_{c}$. For the purpose of obtaining solutions to the wave equation in this geometry, we find it convenient to use a different coordinatization of the radial direction \cite{Das:1999pt}, 
\begin{equation} 
\begin{split}
\mathfrak{z} = \sqrt{f(r)} = \sqrt{1 - \frac{r_{h}^{2}}{r^{2}}} \ \ \implies \ \ \zeta  + \zeta _{c} =\frac{i}{2 \pi } \log \left(\frac{1+ \sqrt{1-\mathfrak{z}^2}}{1-\sqrt{1-\mathfrak{z}^2}}\right).
\end{split}
\end{equation}
The functions that occur in the gradient expansion have a simple form when written in terms of $\sqrt{f}$. It stands to reason that these functions are just expansions in frequency and momenta of a more general function naturally expressed in terms of this new variable. We emphasize here that the coordinate $\mathfrak{z}$ is not a smooth parametrization of AdS$_{3}$ and has a square root branch cut at the future horizon, $\mathfrak{z} = 0$ --- while deriving the infalling solution, we will have to take some care in ensuring that this does not make an undue appearance. Adapted to these new co-ordinates, the $AdS_{3}$ metric is now given by,
\begin{equation} 
\begin{split}
\mathrm{d}s^{2} = - \, r_{h}^{2} \, \frac{\mathfrak{z}^{2}}{1 - \mathfrak{z}^{2}} \; \mathrm{d}v^{2} \, + \, 2 \, r_{h} \, \frac{\mathfrak{z}}{(1-\mathfrak{z}^{2})^{3/2}} \; \mathrm{d}v \, \mathrm{d}\mathfrak{z} \, + \,  r_{h}^{2} \, \frac{1}{1-\mathfrak{z}^{2}} \; \mathrm{d}x^{2}.
\end{split}
\end{equation}
The tetrad frame that we have chosen reads,
\begin{equation} 
\begin{split}
E^{(v)} = r_{h} \, \frac{1 + \mathfrak{z}^{2}}{\sqrt{1-\mathfrak{z}^{2}}} \, \mathrm{d}v - \frac{\mathfrak{z}}{1-\mathfrak{z}^{2}} \, \mathrm{d}\mathfrak{z}, \quad E^{(\mathfrak{z})} = r_{h} \, \sqrt{1 - \mathfrak{z}^{2}} \, \mathrm{d}v - \frac{\mathfrak{z}}{1-\mathfrak{z}^{2}} \, \mathrm{d}\mathfrak{z}, \quad E^{(x)} = r_{h} \, \frac{1}{\sqrt{1-\mathfrak{z}^{2}}} \, \mathrm{d}x.
\end{split}
\end{equation}
Finally, a little algebra shows that the Dirac equation (in Fourier space) takes the following form.
\begin{equation}\label{DEFourier}
\begin{split}
\big(1-\mathfrak{z}^{2}\big) \, \bigg[ \ \mathbb{\Gamma } \, \partial _{\mathfrak{z}} \bigg(\frac{\mathfrak{z} \,  \Psi }{\sqrt{1-\mathfrak{z}^{2}}}  \bigg) + \frac{\mathbb{\Gamma }^{\dagger}}{\mathfrak{z}} \  \partial _{\mathfrak{z}} \bigg(\frac{\Psi }{\sqrt{1-\mathfrak{z}^{2}}}  \bigg)\bigg] + \frac{2 i \omega }{r_{h}} \ \mathbb{\Gamma } \, \Psi  + \frac{i  k}{r_{h}} \ \Gamma ^{(x)} \, \Psi  - \frac{m}{\sqrt{1-\mathfrak{z}^{2}}} \, \Psi  = 0.
\end{split}
\end{equation}
Borrowing from the conventions laid out in Appendix \ref{Spinor and Gamma Conventions}, we use the following explicit representations of the above matrices \eqref{Matrices in d=2},
\begin{equation} 
\begin{split}
\mathbb{\Gamma } = \frac{1}{2} \begin{pmatrix}
1 & 1 \\
-1 & -1
\end{pmatrix} \, , \qquad \mathbb{\Gamma } = \frac{1}{2}\begin{pmatrix}
1 & -1 \\
1 & -1
\end{pmatrix} \, , \qquad \Gamma ^{(x)} = \begin{pmatrix}
0 & 1 \\
1 & 0
\end{pmatrix} .
\end{split}
\end{equation}

Rather that solve this differential equation for the infalling Green's function $S^{\mathrm{in}}(\omega ,\mathfrak{z},k)$ directly, it is useful to design an ansatz that smoothens and streamlines the rest of the presentation. To this end, we choose,
\begin{equation}\label{Infalling Ansatz}
\begin{split}
&S ^{\mathrm{in}}(\omega ,\mathfrak{z} ,k) =  \frac{1}{\xi ^{\mathrm{in}}_{1 \, c}}  \, \frac{(1-\mathfrak{z}^2)^{\frac{1}{2}-\frac{m}{2}}}{r_{h}^{1-m}}  \ \Big(1+\sqrt{1-\mathfrak{z}^2}\Big)^{-\frac{i \omega }{r_{h}}} \ \frac{1}{2} \, \Bigg[ \,  \big(\xi ^{\mathrm{in}} _{1} + \xi ^{ \mathrm{in}}_{2} \big)  \ \mathbb{1}+  \frac{\big(\xi ^{\mathrm{in}} _{1} - \xi ^{ \mathrm{in}}_{2} \big)}{\mathfrak{z}} \ \Gamma ^{(\zeta )} \, \Bigg] \,  S_{0} ,
\\
& \qquad \qquad \qquad \qquad \ \ \xi _{1}^{\mathrm{in}}(\omega , \mathfrak{z} ,k) = \sqrt{1 + \mathfrak{z} } \  \Big( \, \chi _{2}(\omega , \mathfrak{z} ,k) + \mathfrak{z}  \, \chi _{1}(\omega , \mathfrak{z} ,k) \, \Big) ,
\\
& \qquad \qquad \qquad \qquad \ \ \xi _{2}^{\mathrm{in}}(\omega , \mathfrak{z} ,k) = \sqrt{1 - \mathfrak{z} } \  \Big( \, \chi _{2}(\omega , \mathfrak{z} ,k) - \mathfrak{z}  \, \chi _{1}(\omega , \mathfrak{z} ,k) \, \Big) .
\end{split}
\end{equation}
Here $\xi ^{\mathrm{in}}_{1 \, c} \equiv \xi ^{\mathrm{in}}_{1} (\omega ,0,k)$, ensures that this solutions is properly normalized. This ansatz deserves some explanation. Recall the structure of the gradient expansion of the massive infalling solution \eqref{Massive ingoing solution first order}. The functions fell into two classes: the `$+$' functions were regular, with a constant sign across the future horizon and always accompanied the indentity matrix. The`$-$' functions flipped sign at $\sqrt{f} = 0$. This latter set of functions always occured with a multiplicative factor of $\sqrt{f}^{-1}$ and appeared as coefficients of $\Gamma ^{(v)}$. Finally, all of this multiplied the matrix $S_{0}$. It is this structure that we have explicitly incorporated into the construction of \eqref{Infalling Ansatz}. Indeed, we will soon show that the functions $\chi _{1}$ and $\chi _{2}$ admit solutions that are regular at the horizon. Assuming this for now, we have,
\begin{equation} 
\begin{split}
\xi^{\mathrm{in}} _{1} +  \xi ^{\mathrm{in}}_{2} =  \kappa (\mathfrak{z} ) \; \chi _{2} +  \frac{\mathfrak{z}^{2}}{\kappa (\mathfrak{z})}  \;  \chi _{1} , \qquad \xi^{\mathrm{in}} _{1} -  \xi ^{\mathrm{in}}_{2} = \mathfrak{z} \, \Big( \, \frac{1}{\kappa (\mathfrak{z} )} \; \chi _{2} + \kappa (\mathfrak{z} ) \;  \chi _{1}  \Big),
\end{split}
\end{equation} 
where $\kappa (\mathfrak{z} ) = (1+\mathfrak{z})^{1/2} + (1-\mathfrak{z})^{1/2}$. Now the functions $\kappa (\mathfrak{z})$, $\kappa (\mathfrak{z})^{-1}$ and $\mathfrak{z}^{2}$ are all regular at the future horizon. Thus, the combination $(\xi ^{\mathrm{in}}_{1} + \xi ^{\mathrm{in}}_{2})$ is indeed a `$+$' function and the combination $(\xi ^{\mathrm{in}}_{1} - \xi ^{\mathrm{in}}_{2})$ is a `$-$' function, as claimed. Finally, the first two terms, being proportional to $r^{m-1}$, facilitate the discussion of the normalization. The third term is a convenient factor that simplifies the resulting differential equations. 

Note, with our choice of basis, the constant boundary--to--bulk matrix $S_{0}$ is given by \eqref{Matrices in d=2},
\begin{equation} 
\begin{split}
S_{0} = \begin{pmatrix}
1 \\ 
-1
\end{pmatrix}.
\end{split}
\end{equation}
We emphasize here that $\chi _{1,2}$ are employed only as a crutch to obtain infalling solutions to the Dirac equation and shed light on its analytical structure. Once this done, repackaging these functions in terms of $\xi ^{\mathrm{in}}_{1}$ and $\xi ^{\mathrm{in}}_{2}$ tidies up the rest of the analysis. The differential equations for $\chi _{1}$ and $\chi _{2}$ obtained by substituting the ansatz \eqref{Infalling Ansatz} into the Dirac equation \eqref{DEFourier} are,
\begin{equation}\label{Chi 1 2 ODE}
\begin{split}
&\frac{(1+\mathfrak{z})}{2} \, \left( \chi _2 ' + \mathfrak{z} \, \chi _1  '\right) + \big(\mathfrak{p}_{-}-2 \overline{h} \big) \, \chi _2 + \big(1+\mathfrak{p}_{+}-2 \overline{h} \big)\,  \mathfrak{z} \, \chi _1 +  \big(\mathfrak{p}_{+}+\mathfrak{p}_{-}-2 \overline{h} \big) \, \chi _1 = 0 ,
\\
&\frac{(1-\mathfrak{z})}{2} \, \left( \chi _2 ' - \mathfrak{z} \, \chi _1 '\right) - \big(\mathfrak{p}_{-}-2 \overline{h} \big) \, \chi _2 + \big(1+\mathfrak{p}_{+}-2 \overline{h} \big) \,  \mathfrak{z} \, \chi _1 -  \big(\mathfrak{p}_{+}+\mathfrak{p}_{-}-2 \overline{h} \big) \, \chi _1 = 0 .
\end{split}
\end{equation}
Here prime denotes differentiation w.r.t. $\mathfrak{z}$. To keep expressions compact, we have introduced the following linear combinations of the parameters that appear in the solutions presented below,
\begin{equation} 
\begin{split}
&\mathfrak{p}_{+} = \overline{h} +  \frac{i \,\beta }{4\pi } \, (k - \omega) , \qquad \mathfrak{p}_{-} = h  - \frac{i \, \beta }{4 \pi } \, (k + \omega).
\end{split}
\end{equation}
Here, $h + \overline{h}= \Delta$ and $h - \overline{h} = s$. Recall that in $d=2$, $\Delta  = m+1$ and $s = \frac{1}{2}$ characterize the dual fermionic operator of the boundary conformal field theory ($2 \overline{h}  = \Delta  - s$  is the twist of this operator). Beginning with \eqref{Chi 1 2 ODE}, it is easy to obtain decoupled, second order ODEs for $\chi _{1}$ and $\chi _{2}$. As promised, these differential equations have solutions that are smooth at the future horizon, given by \footnote{In presenting the solutions, we have used the Euler transformation, \[{}_2\tilde{F}_1(a,b;c;z) = (1 -z)^{c - a - b} \, {}_{2}\tilde{F}_{1}(c-a,c-b;c;z).\] We observe that the solutions $\chi _{1}$ and $\chi _{2}$ are not independent, but are related via,
	\begin{equation*} 
	\begin{split}
	\frac{\partial }{\partial \mathfrak{z} }\Big[\left(1-\mathfrak{z} ^2\right)^{\mathfrak{p}_- -2 \overline{h} } \chi _{2}(\omega ,\mathfrak{z} ,k)\Big] = 2 \mathfrak{z} \left(\mathfrak{p}_--1\right) \left(1-\mathfrak{z} ^2\right)^{-1 +\mathfrak{p}_--2 \overline{h} }\chi _{1}(\omega ,\mathfrak{z} ,k).
	\end{split}
	\end{equation*}},
\begin{equation}\label{Infalling Chi  Solutions}
\begin{split}
&\chi _{1}(\omega ,\mathfrak{z} ,k) = \left(2 \overline{h} - \mathfrak{p}_- \right)  \left(1-\mathfrak{z} ^2\right)^{2 \overline{h} -1} \, _2\tilde{F}_1\left(\mathfrak{p}_+,\mathfrak{p}_-;1 + \mathfrak{p}_+ +\mathfrak{p}_- - 2 \overline{h} ;\mathfrak{z} ^2\right) ,
\\
&\chi _{2}(\omega ,\mathfrak{z} ,k) = \left(1-\mathfrak{z} ^2\right)^{2 \overline{h} -1}  \, _2\tilde{F}_1\left(\mathfrak{p}_+,\mathfrak{p}_--1; \mathfrak{p}_+ +\mathfrak{p}_- - 2 \overline{h} ;\mathfrak{z} ^2\right) .
\end{split}
\end{equation}

Putting this all together, the normalized infalling boundary--to--bulk Green's function reads,
\begin{equation}\label{2D Infalling Green's Function}
\begin{split}
S ^{\mathrm{in}}(\omega , \zeta ,k) &= \frac{1}{\xi ^{\mathrm{in}}_{1 \, c}} \ \bigg(\frac{2\pi }{\beta }\bigg)^{\Delta -2} \Bigg(\frac{1 - e^{2\pi i (\zeta + \zeta _{c})}}{1 + e^{2\pi i (\zeta + \zeta _{c})}}\Bigg)^{2-\Delta }  \Bigg(\frac{1 + e^{2\pi i (\zeta + \zeta _{c})}}{1 + e^{2\pi i  \zeta _{c}}}\Bigg) ^{\Delta - \mathfrak{p}_{+} - \mathfrak{p}_{-}}
\\
& \qquad \qquad \qquad \qquad \quad \times  \, \frac{1}{2} \ \Bigg[  \Big(\xi ^{\mathrm{in}} _{1} + \xi ^{ \mathrm{in}}_{2} \Big)  \, \mathbb{1}+  \cos \pi (\zeta +\zeta _{c}) \; \Big(\xi ^{\mathrm{in}} _{1} - \xi ^{ \mathrm{in}}_{2} \Big) \, \Gamma ^{(\zeta )}\Bigg] \,  S_{0},
\end{split}
\end{equation}
where,
\begin{equation}\label{2D Infalling Functions}
\begin{split}
&\xi ^{\mathrm{in}}_{1}  =  \Bigg(\frac{1 - e^{2\pi i (\zeta + \zeta _{c})}}{1 + e^{2\pi i (\zeta + \zeta _{c})}}\Bigg)^{4 \overline{h} - 2} \Big(1 + \sec \pi  (\zeta  + \zeta _{c})\Big)^{\frac{1}{2}}
\\
& \qquad \qquad \times   \Bigg[  {}_2\tilde{F}_1\Big(\mathfrak{p}_+,\mathfrak{p}_--1; \mathfrak{p}_+ +\mathfrak{p}_- -2 \overline{h} ; \; \sec ^{2}\pi  (\zeta  + \zeta _{c})\Big)
\\
&\qquad \qquad \qquad \quad \ \, + \big(2 \overline{h} - \mathfrak{p}_- \big) \, \sec \pi  (\zeta  + \zeta _{c})  \ _2\tilde{F}_1\Big(\mathfrak{p}_+,\mathfrak{p}_-;1 + \mathfrak{p}_+ +\mathfrak{p}_- - 2 \overline{h} ; \; \sec ^{2}\pi  (\zeta  + \zeta _{c})\Big)\Bigg],
\\
&\xi ^{\mathrm{in}}_{2}  =  \Bigg(\frac{1 - e^{2\pi i (\zeta + \zeta _{c})}}{1 + e^{2\pi i (\zeta + \zeta _{c})}}\Bigg)^{4 \overline{h} - 2} \Big(1 - \sec \pi  (\zeta  + \zeta _{c})\Big)^{\frac{1}{2}}
\\
& \qquad \qquad \times   \Bigg[  {}_2\tilde{F}_1\Big(\mathfrak{p}_+,\mathfrak{p}_--1; \mathfrak{p}_+ +\mathfrak{p}_- -2 \overline{h} ; \; \sec ^{2}\pi  (\zeta  + \zeta _{c})\Big)
\\
&\qquad \qquad \qquad \quad \ \, - \big(2 \overline{h} - \mathfrak{p}_- \big) \, \sec \pi  (\zeta  + \zeta _{c})  \ _2\tilde{F}_1\Big(\mathfrak{p}_+,\mathfrak{p}_-;1 + \mathfrak{p}_+ +\mathfrak{p}_- - 2 \overline{h} ; \; \sec ^{2}\pi  (\zeta  + \zeta _{c})\Big)\Bigg],
\\
& \xi ^{\mathrm{in}}_{1 \, c} \equiv \xi ^{\mathrm{in}}_{1}(\omega , 0 ,k)  = \frac{\pi }{\sqrt{2}} \; \frac{ \csc (2 \overline{h} \, \pi )}{ \Gamma \left(1 - 2 \overline{h}  \right) \, \Gamma \left(\mathfrak{p}_+\right) \, \Gamma \left(\mathfrak{p}_-\right)}.
\end{split}
\end{equation}

We have written out this Green's function in terms of the mock tortoise coordinate $\zeta $, and determined the normalization constant $\xi ^{\mathrm{in}}_{1 \, c}$.  This solution is smooth at the future horizon --- while this is a subtle point,  our discussion preceding \eqref{Chi 1 2 ODE} guarantees it. Thus, \eqref{2D Infalling Green's Function} can be safely elevated to an infalling boundary--to--bulk Green's function on the grSK spacetime that is smooth around the horizon cap.

The outgoing boundary--to--bulk Green's function can now be immediately worked out using the `time--reversing' map detailed in $\S$\ref{Subsection: Time Reversing Solutions}. This yields,  
\begin{equation}\label{2D Outgoing Green's Function}
\begin{split}
S  ^{\mathrm{out}}(\omega , \zeta  ,k) &= \mathfrak{T} \cdot S ^{\mathrm{in}}(-\omega , \zeta , -k) \, e^{- \beta \omega \zeta }
\\
&= \frac{e^{-\beta \omega \zeta }}{\xi ^{\mathrm{rev}}_{1 \, c}} \ \bigg(\frac{2\pi }{\beta }\bigg)^{\Delta -2} \ \Bigg(\frac{1 - e^{2\pi i (\zeta + \zeta _{c})}}{1 + e^{2\pi i (\zeta + \zeta _{c})}}\Bigg)^{2-\Delta }  \Bigg(\frac{1 + e^{2\pi i (\zeta + \zeta _{c})}}{1 + e^{2\pi i  \zeta _{c}}}\Bigg) ^{-\Delta + \mathfrak{p}_{+} + \mathfrak{p}_{-}}
\\
& \qquad \qquad \qquad \qquad \times  \, \frac{1}{2} \ \Bigg[  \Big(\xi^{\mathrm{rev}} _{1} - \xi ^{ \mathrm{rev}}_{2} \Big)  \, \mathbb{1}+  \cos \pi (\zeta +\zeta _{c}) \; \Big(\xi^{\mathrm{rev}} _{1} + \xi ^{ \mathrm{rev}}_{2} \Big) \, \Gamma ^{(\zeta )}\Bigg] \,  S_{0},
\end{split}
\end{equation}
where we have defined\footnote{In terms of the new parameters $(\mathfrak{p}_{+}, \mathfrak{p}_{-})$, the involution $(\omega ,k) \rightarrow (-\omega,-k) $ reads,
	\[\mathfrak{p}_{+} \rightarrow  2 \overline{h} - \mathfrak{p}_{+},  \qquad \mathfrak{p}_{-}\rightarrow 2 h  - \mathfrak{p}_{-}.\]},
\begin{equation} 
\begin{split}
&\xi ^{\mathrm{rev}}_{j}(\omega , \zeta ,k) \equiv \xi ^{\mathrm{in}}_{j}(-\omega , \zeta ,-k), \quad \ j \in \{1,2\}.
\end{split}
\end{equation}
Writing these out explicitly,
\begin{equation} 
\begin{split}
&\xi ^{\mathrm{rev}}_{1} = \Bigg(\frac{1 - e^{2\pi i (\zeta + \zeta _{c})}}{1 + e^{2\pi i (\zeta + \zeta _{c})}}\Bigg)^{4 \overline{h} - 2} \big(1 + \sec \pi  (\zeta  + \zeta _{c})\big)^{\frac{1}{2}} 
\\
& \qquad \quad  \times \Bigg[ {}_2\tilde{F}_1\Big(2 \overline{h} -\mathfrak{p}_+,2 \overline{h} -\mathfrak{p}_-;1 - \mathfrak{p}_+ - \mathfrak{p}_- + 2 \overline{h} ; \ \sec^{2} \pi  (\zeta  + \zeta _{c})\Big)
\\
& \qquad \qquad \quad + \sec \pi  (\zeta  + \zeta _{c}) \left(\mathfrak{p}_- - 1\right) \, _2\tilde{F}_1\Big(2 \overline{h} -\mathfrak{p}_+,2 h -\mathfrak{p}_-; 2 - \mathfrak{p}_+ - \mathfrak{p}_- + 2 \overline{h} ; \ \sec^{2} \pi  (\zeta  + \zeta _{c})\Big)\Bigg] ,
\\
&\xi ^{\mathrm{rev}}_{2} = \Bigg(\frac{1 - e^{2\pi i (\zeta + \zeta _{c})}}{1 + e^{2\pi i (\zeta + \zeta _{c})}}\Bigg)^{4 \overline{h} - 2} \big(1 - \sec \pi  (\zeta  + \zeta _{c})\big)^{\frac{1}{2}} 
\\
& \qquad \quad  \times \Bigg[ {}_2\tilde{F}_1\Big(2 \overline{h} -\mathfrak{p}_+,2 \overline{h} -\mathfrak{p}_-;1 - \mathfrak{p}_+ - \mathfrak{p}_- + 2 \overline{h} ; \ \sec^{2} \pi  (\zeta  + \zeta _{c})\Big)
\\
& \qquad \qquad \quad - \sec \pi  (\zeta  + \zeta _{c}) \left(\mathfrak{p}_- - 1\right) \, _2\tilde{F}_1\Big(2 \overline{h} -\mathfrak{p}_+, 2 h -\mathfrak{p}_-; 2 - \mathfrak{p}_+ - \mathfrak{p}_- + 2 \overline{h} ; \ \sec^{2} \pi  (\zeta  + \zeta _{c})\Big)\Bigg] ,
\\
& \xi ^{\mathrm{rev}}_{1 \, c} = \frac{\pi }{\sqrt{2}} \; \frac{ \csc (2 \overline{h} \, \pi )}{ \Gamma \left(1 - 2 \overline{h} \right) \, \Gamma \left(2 \overline{h} - \mathfrak{p}_+\right) \, \Gamma \left(2 h  - \mathfrak{p}_-\right)}.
\end{split}
\end{equation}
As anticipated, the outgoing boundary--to--bulk Green's function is not regular around the horizon cap, with branch cuts at $\sqrt{f} = 0$ arising from the factor $e^{-\beta \omega \zeta }$ and the matrix $\mathfrak{T}$.

Finally, the full solution $\Psi $ is obtained by combining the two boundary--to--bulk  Green's functions and sourcing them at the two boundaries $\zeta  = 0$ and $\zeta =  1$ by $\psi _{L}$ and $\psi _{R}$ respectively. Following the conventions laid out in Appendix \ref{Spinor and Gamma Conventions}, these are single component Weyl spinors (i.e. ordinary complex functions) in $d=2$. Expressing them in the \emph{average--difference} basis \eqref{ad Basis}, the full solution reads,
\begin{equation} 
\begin{split}
\Psi (\omega , \zeta  , k) = S ^{\mathrm{in}}(\omega , \zeta , k) \ \Big[\psi _{a} + \Big(\frac{1}{2} -n_{\omega }^{\mathrm{FD}}\Big) \,  \psi _{d}\Big] - S ^{\mathrm{out}} (\omega , \zeta , k) \ \Big[\big(1-n^{\mathrm{FD}}_{\omega }\big) \, \psi _{d}\Big]
\end{split}
\end{equation}

\subsection{The influence phase}
We calculate the influence phase for the fermionic probe field by evaluating the Dirac action on--shell. This reduces to a boundary term, of which we select only the finite part
\begin{equation}\label{2D IF Structure}
\begin{split}
\mathds{S}_{\mathrm{IF}} &= i \int_{\omega ,k} \  r_{c}^{2} \; \bigg\{ \overline{\Psi} (\omega , \zeta  ,k) \  \mathcal{P}^{\zeta }_{-} \, \Psi (\omega , \zeta ,k)\bigg\}_{ r_{c}^{-2}} \ \Bigg|^{\zeta =1}_{\zeta =0} = i \int_{\omega ,k} \  r_{c}^{2} \ \overline{\mathcal{P}_{+}^{\zeta } \; \Psi}  \ \bigg\{ \mathcal{P}^{\zeta }_{-} \, \Psi \bigg\}_{r_{c}^{-\Delta }}  \ \Bigg|^{\zeta =1}_{\zeta =0}.
\end{split}
\end{equation}
Once again, the notation asks us to select the term from the expression enclosed by the curly brackets with the fall-off given by the subscript. For the ease of computation, we calculate the contributions from the infalling and outgoing Green's function separately. 

Let us begin with the infalling solution. The two constituent functions of the infalling boundary--to--bulk Green's function, $\xi ^{\mathrm{in}}_{1}$ and $\xi ^{\mathrm{in}}_{2}$, effectively exchange their roles across the horizon cap, i.e.,
\begin{equation} 
\begin{split}
\lim _{\zeta  \rightarrow 0} \ \xi ^{\mathrm{in}}_{1} = \lim _{\zeta  \rightarrow 1} \ \xi ^{\mathrm{in}}_{2} \, , \qquad \qquad \lim _{\zeta  \rightarrow 0} \ \xi ^{\mathrm{in}}_{2} = \lim _{\zeta  \rightarrow 1} \ \xi ^{\mathrm{in}}_{1}.
\end{split}
\end{equation} 
The reason for this can be traced to the $\mathfrak{z}=\sqrt{f}$ in their definitions \eqref{2D Infalling Functions}, which picks up a minus sign across the future horizon. As a consequence, the boundary values of the infalling boundary--to--bulk Green's function \eqref{2D Infalling Green's Function} can be compactly written as,
\begin{equation} 
\begin{split}
&\lim_{\zeta \rightarrow 0} \ S^{\mathrm{in}} (\omega ,\zeta ,k) =  r_{c}^{\Delta -2 } \  \Big[ \,  \mathcal{P}_{+}^{\zeta } + \Xi^{\mathrm{in}}  \ \mathcal{P}_{-}^{\zeta } \, \Big] \,  S_{0} ,
\\
&\lim_{\zeta \rightarrow 1} \ S^{\mathrm{in}}(\omega , \zeta , k) = r_{c}^{\Delta -2 } \ \Big[ \, \mathcal{P}_{+}^{\zeta } + \Xi^{\mathrm{in}}  \ \mathcal{P}_{-}^{\zeta } \,  \Big] \,  S_{0} ,
\end{split}
\end{equation}   
where  we have defined $\Xi^{\mathrm{in}} = (\xi ^{\mathrm{in}}_{1 \, c})^{-1} \lim _{\zeta \rightarrow 0} \  \xi ^{\mathrm{in}}_{2}$. Thus, the contribution from this Green's function to the integrand of the influence phase is,
\begin{equation} 
\begin{split}
&\lim_{\zeta \rightarrow 0} \  r_{c}^{\Delta } \, \Big\{ \mathcal{P}^{\zeta }_{-} \ S^{\mathrm{in}} (\omega ,\zeta ,k) \Big\}_{r_{c}^{-\Delta }} =  \ r_{c}^{2\Delta - 2} \ \Big\{ \Xi^{\mathrm{in}} \Big\}_{r_{c}^{2-2\Delta }}  \ \mathcal{P}_{-}^{\zeta } \,  S_{0}
\\
&\lim_{\zeta \rightarrow 1} \ r_{c}^{\Delta } \, \Big\{ \mathcal{P}^{\zeta }_{-} \ S^{\mathrm{in}} (\omega ,\zeta ,k) \Big\}_{r_{c}^{-\Delta }} = \ r_{c}^{2\Delta - 2} \ \Big\{ \Xi^{\mathrm{in}} \Big\}_{r_{c}^{2-2\Delta }}  \ \mathcal{P}_{-}^{\zeta } \,  S_{0},
\end{split}
\end{equation}
with,
\begin{equation}\label{Xi in}
\begin{split}
r_{c}^{2\Delta -2} \; \Big\{ \Xi ^{\mathrm{in}} \Big\}_{ r_{c}^{2-2\Delta }} = r_{h}^{2\Delta -2} \, \frac{ \Gamma \left(1 - 2 \overline{h} \right) \, \Gamma \left(\mathfrak{p}_+\right) \, \Gamma \left(\mathfrak{p}_-\right)}{ \Gamma \left(2 \overline{h} \right)  \, \Gamma \left(\mathfrak{p}_+ - 2 h\right) \, \Gamma \left( \mathfrak{p}_- - 2 \overline{h} \right)}.
\end{split}
\end{equation}

In a similar vein, using \eqref{2D Outgoing Green's Function}, the contribution from the outgoing boundary--to--bulk Green's function to the integrand of \eqref{2D IF Structure} can be expressed as,
\begin{equation}\label{}
\begin{split}
&\lim_{\zeta \rightarrow 0} \  r_{c}^{\Delta } \, \Big\{ \mathcal{P}^{\zeta }_{-} \ S^{\mathrm{out}} (\omega ,\zeta ,k) \Big\}_{r_{c}^{-\Delta }} =  - \, r_{c}^{2\Delta -2} \; \Big\{ \Xi ^{\mathrm{rev}} \Big\}_{ r_{c}^{2-2\Delta }}  \ \mathcal{P}_{-}^{\zeta } \,  S_{0} ,
\\
&\lim_{\zeta \rightarrow 0} \  r_{c}^{\Delta } \, \Big\{ \mathcal{P}^{\zeta }_{-} \ S^{\mathrm{out}} (\omega ,\zeta ,k) \Big\}_{r_{c}^{-\Delta }} =   e^{-\beta \omega} \, r_{c}^{2\Delta -2} \; \Big\{ \Xi ^{\mathrm{rev}} \Big\}_{ r_{c}^{2-2\Delta }}  \ \mathcal{P}_{-}^{\zeta } \,  S_{0},
\end{split}
\end{equation}
where we have defined an analogous ratio $\Xi^{\mathrm{rev}} = (\xi ^{\mathrm{rev}}_{1 \, c})^{-1} \lim _{\zeta \rightarrow 0} \  \xi ^{\mathrm{rev}}_{2}$. The value of this asymptotic ratio can be immediately read off by simply reversing the frequency and momenta in  \eqref{Xi in}, giving,
\begin{equation} 
\begin{split}
r_{h}^{2\Delta -2} \; \Big\{ \Xi ^{\mathrm{rev}} \Big\}_{ r_{c}^{2-2\Delta }} = r_{h}^{2\Delta -2} \, \frac{ \Gamma \left(1-2 \overline{h}  \right) \, \Gamma \left(2 \overline{h} - \mathfrak{p}_+\right) \, \Gamma \left(2 h  - \mathfrak{p}_-\right)}{\Gamma \left( 2 \overline{h} \right) \,  \Gamma \left( 1 - \mathfrak{p}_-  \right) \, \Gamma \left(1 - \mathfrak{p}_+ \right)} .
\end{split}
\end{equation}
Equivalently, because the frequency and momenta always occur with a coefficient of $i$ in these expressions, the two asymptotic ratios  are related to one another by complex conjugation.
\begin{equation}\label{Complex conjugation}
\begin{split}
\Big\{ \Xi ^{\mathrm{in}} \Big\}_{ r_{c}^{2-2\Delta }}^{*} = \Big\{ \Xi ^{\mathrm{rev}} \Big\}_{ r_{c}^{2-2\Delta }}
\end{split}
\end{equation}

Putting this together, the integrand of the influence phase \eqref{2D IF Structure} evaluated in the average--difference basis at the two boundaries is,
\begin{equation}\label{2D IF Boundary Contributions}
\begin{split}
& i \, r_{c}^{2} \, \Big\{ \overline{\Psi } \mathcal{P}^{\zeta }_{-} \Psi \Big\}_{ r_{c}^{-2}} \bigg|_{\zeta  = 0} 
\\
& \qquad \qquad = - i \, r_{c}^{2\Delta -2} \,  \Big(\psi _{a}^{*} - \frac{1}{2}\psi _{d}^{*}\Big)\, \bigg[ \Big(\psi _{a} +  \Big(\frac{1}{2}-n^{\mathrm{FD}}_{\omega }\Big) \, \psi _{d}\Big) \; \big\{ \Xi ^{\mathrm{in}} \big\} + \big(1-n^{\mathrm{FD}}_{\omega }\big) \, \psi _{d} \; \big\{ \Xi ^{\mathrm{rev}} \big\} \bigg] ,
\\
& i \, r_{c}^{2} \, \Big\{ \overline{\Psi }  \mathcal{P}^{\zeta }_{-}  \Psi \Big\}_{ r_{c}^{-2}} \bigg|_{\zeta  = 1}
\\ 
& \qquad \qquad = - i \, r_{c}^{2\Delta -2}  \Big(\psi _{a}^{*} + \frac{1}{2}\psi _{d}^{*}\Big)\, \bigg[ \Big(\psi _{a} + \Big(\frac{1}{2}-n^{\mathrm{FD}}_{\omega }\Big) \, \psi _{d}\Big) \; \big\{ \Xi ^{\mathrm{in}} \big\} - \,  n_{\omega }^{\mathrm{FD}} \, \psi _{d} \; \big\{ \Xi ^{\mathrm{rev}} \big\} \bigg].
\end{split}
\end{equation}
where we have adopted the shorthand $\big\{\Xi ^{\mathrm{in} / \mathrm{rev}}\big\} = \big\{\Xi ^{\mathrm{in} / \mathrm{rev}}\big\}_{r_{c}^{2 - 2\Delta }}$. Evaluating the difference of the above two contributions, the full Schwinger--Keldysh influence phase can be neatly written down in the form,
\begin{equation}\label{2D IF}
\begin{split}
\mathds{S}_{\mathrm{IF}} = \int \frac{\mathrm{d}{\omega } \ \mathrm{d}{k}}{(2\pi )^{2}} \   \Big(\mathfrak{I}_{ad} \  \psi _{d}^{*} \, \psi _{a} +  \mathfrak{I}_{da} \ \psi _{a}^{*} \, \psi _{d}  + \frac{1}{2} \tanh \Big(\frac{\beta \omega }{2}\Big) \, \big(\mathfrak{I}_{ad} - \mathfrak{I}_{da}\big) \ \psi _{d}^{*} \, \psi _{d} \Big) ,
\end{split}
\end{equation} 
where,
\begin{equation}\label{2D IF Functions}
\begin{split}
&\mathfrak{I}_{ad} (\omega ,k) = - i \, r_{c}^{2\Delta -2} \; \Big\{ \Xi ^{\mathrm{in}} \Big\}_{ r_{c}^{2-2\Delta }} =- i \, \frac{ \Gamma \left(1 - 2 \overline{h} \right) \, \Gamma \left(\mathfrak{p}_+\right) \, \Gamma \left(\mathfrak{p}_-\right)}{ \Gamma \left(2 \overline{h} \right) \,  \Gamma \left( \mathfrak{p}_- - 2 \overline{h} \right) \, \Gamma \left(\mathfrak{p}_+ - 2 h\right)} \, r_{h}^{2\Delta -2} ,
\\
&\mathfrak{I}_{da} (\omega ,k) = i \, r_{c}^{2\Delta -2} \; \Big\{ \Xi ^{\mathrm{rev}} \Big\}_{ r_{c}^{2-2\Delta }} = i \, \frac{ \Gamma \left(1-2 \overline{h}  \right) \, \Gamma \left(2 \overline{h} - \mathfrak{p}_+\right) \, \Gamma \left(2h  - \mathfrak{p}_-\right)}{\Gamma \left( 2 \overline{h} \right) \,  \Gamma \left( 1 - \mathfrak{p}_-  \right) \, \Gamma \left(1 - \mathfrak{p}_+ \right)} \, r_{h}^{2\Delta -2} .
\end{split}
\end{equation}
Because of \eqref{Complex conjugation}, $\mathfrak{I}_{ad}(\omega ,k)^{*} = \mathfrak{I}_{da}(\omega , k)$.

A few comments are in order. Let us confine our attention to the function $\mathfrak{I}_{ad}(\omega ,k)$ defined above. Writing this in terms of the bulk variables $(m,\omega ,k)$, we have,
\begin{equation}\label{2D Retarded Green's Funciton}
\begin{split}
\mathfrak{I}_{ad} (\omega ,k) =- i \, \frac{ \Gamma \left(\frac{1}{2} - m \right) \, \Gamma \left(\frac{1}{4} + \frac{m}{2} +  \frac{i \,\beta }{4\pi } \, (k - \omega) \right) \, \Gamma \left(\frac{3}{4} + \frac{m}{2}  - \frac{i \, \beta }{4 \pi } \, (k + \omega)\right)}{ \Gamma \left(\frac{1}{2} + m \right) \,  \Gamma \left(\frac{1}{4} - \frac{m}{2}  - \frac{i \, \beta }{4 \pi } \, (k + \omega)\right) \, \Gamma \left(\frac{3}{4} - \frac{m}{2} +  \frac{i \,\beta }{4\pi } \, (k - \omega)\right)} \, r_{h}^{2\Delta -2} .
\end{split}
\end{equation}
This is just the retarded Green's function $\mathcal{S}^{\mathrm{ret}} (\omega ,k)$ for a 2$d$ fermionic CFT on the real line. Indeed, this quantity can be obtained directly, and is a result quoted extensively in the literature. This Green's function has poles whenever the argument of the functions in the numerator take on negative integer values. This leads to two denumerable sets of poles, given by,
\begin{equation} 
\begin{split}
&\omega_{n} = + k -4\pi i \, T \, (\overline{h}  + n ) ,
\\
&\omega_{n} =- k -4\pi i \,  T \, (h +n), \qquad n \in \mathds{Z}^{+}_{0}
\end{split}
\end{equation} 
These poles appear only in the lower half of the complex plane, consistent with what we expect for a retarded Green's function. As pointed out in \cite{Birmingham:2001pj}, they precisely coincide with the quasi--normal modes of a BTZ blackhole . The influence phase also contains information about the advanced Green's function $\mathcal{S}^{\mathrm{adv}}(\omega ,k)$. Using \eqref{Complex conjugation} and \eqref{2D IF Functions}, we have,
\begin{equation} 
\begin{split}
\mathcal{S}^{\mathrm{adv}} (\omega ,k) = \mathcal{S}^{\mathrm{ret}}(\omega ,k)^{*} = \mathfrak{I}_{ad}^{*}(\omega ,k) = \mathfrak{I}_{da}(\omega ,k) ,
\end{split}
\end{equation}
where,
\begin{equation} 
\begin{split}
\mathcal{S}^{\mathrm{adv}}(\omega ,k) = i \, \frac{ \Gamma \left(\frac{1}{2} - m \right) \, \Gamma \left(\frac{1}{4} + \frac{m}{2} -  \frac{i \,\beta }{4\pi } \, (k - \omega) \right) \, \Gamma \left(\frac{3}{4} + \frac{m}{2}  + \frac{i \, \beta }{4 \pi } \, (k + \omega)\right)}{ \Gamma \left(\frac{1}{2} + m \right) \,  \Gamma \left(\frac{1}{4} - \frac{m}{2}  + \frac{i \, \beta }{4 \pi } \, (k + \omega)\right) \, \Gamma \left(\frac{3}{4} - \frac{m}{2} - \frac{i \,\beta }{4\pi } \, (k - \omega)\right)} \, r_{h}^{2\Delta -2} .
\end{split}
\end{equation}

Finally, comparing the terms multiplying $\psi _{d}^{*} \, \psi _{d}$ in the integrand of the general Schwinger--Keldysh influence phase with the exact expression we have obtained \eqref{2D IF}, we are lead to the fluctuation--dissipation theorem for this theory.
\begin{equation} 
\begin{split}
\mathfrak{I}_{aa}(\omega ,k) &=  \frac{1}{2} \tanh \Big(\frac{\beta \omega }{2}\Big) \, \big(\mathfrak{I}_{ad} (\omega ,k) - \mathfrak{I}_{da} (\omega ,k)\big) 
\\
&= i \, \tanh \Big(\frac{\beta \omega }{2}\Big) \, \mathrm{Im} \, \big(\mathfrak{I}_{ad} (\omega ,k)\big)
\end{split}
\end{equation}

\subsection{The gradient expansion}
The solutions described in the previous subsections are exact, and therefore valid to all orders in gradient expansion. We now demonstrate how they can be expanded to produce results consistent with those described in the main text, focussing on the infalling boundary--to--bulk Green's function.

\subsection*{Zeroth order} 
At zeroth order in gradient expansion, the functions $\xi ^{\mathrm{in}}_{1,2}$ can be immediately evaluated to give,
\begin{equation} 
\begin{split}
\xi ^{\mathrm{in}}_{1}(0,\zeta ,0) = \frac{\big(1 + \sqrt{f}\big)^{m}}{\sqrt{\pi }} = \frac{1}{\sqrt{\pi }} \, \frac{r_{h}^{m}}{r^{m}}  \ \mathcal{F}, \qquad   \xi ^{\mathrm{in}}_{2}(0,\zeta ,0) = \frac{\big(1 - \sqrt{f}\big)^{m}}{\sqrt{\pi }} = \frac{1}{\sqrt{\pi }} \, \frac{r_{h}^{m}}{r^{m}} \ \frac{1}{\mathcal{F}},
\end{split}
\end{equation}
where $\mathcal{F}(\zeta ) = (1 + \sqrt{f})^{\frac{m}{2}}(1-\sqrt{f})^{-\frac{m}{2}}$ is the function that we had introduced in the main text of this paper while discussing the massive gradient expansion \eqref{F definition}. In particular, at this order in expansion, the normalization $\xi ^{\mathrm{in}}_{1 \, c}$ becomes $\xi ^{\mathrm{in}}_{1 \, c} (0,\zeta ,0)= 2^{m}/\sqrt{\pi }$. Thus, recombining the above functions into \eqref{2D Infalling Green's Function}, we get,
\begin{equation} 
\begin{split}
S^{\mathrm{in}} (0, \zeta , 0) = \frac{1}{r} \, \frac{r_{h}^{m}}{2^{m}}  \, \bigg[ \, \mathcal{F}_+ \, \mathds{1}+\mathcal{F}_- \, \frac{\Gamma ^{(\zeta )}}{\sqrt{f}} \, \Bigg] S_0.
\end{split}
\end{equation}

\subsection*{Higher orders} 
The contributions at higher orders in gradient expansion are determined using the series expansion of the regularized hypergeometric functions.
\begin{equation} 
\begin{split}
_2 \tilde{F}_1 (a,b;c;z) = \sum_{n=0}^{\infty} \frac{(a)_{n} \, (b)_{n}}{\Gamma (c+n)} \, \frac{z^{n}}{n!}
\end{split}
\end{equation}
The parameters $(a, b, c)$ can now be expanded in $(\omega ,k)$ to the desired order. Notice that in order to produce the correct result, we have to resum the contributions at the relevant order from each term in the infinite series.  As an example, we show how this works for the regularized hypergeometric functions $\chi _{1,2}$ tha appear in the infalling solutions at first order in gradient expansion.  From \eqref{Infalling Chi  Solutions}, 
\begin{equation}\label{2D Hypergeometric Expansions}
\begin{split}
&\frac{\partial {\chi _{1}}}{\partial {\omega }} \Bigg|_{0} = \frac{1}{\sqrt{\pi }} \, \frac{i \, \beta }{2\pi } \, \sum_{n=0}^{\infty} \, \frac{\sec^{2n}\pi (\zeta +\zeta _{c})}{ \Gamma (2n + 2)}\left(\prod _{l=0}^{2 n} \big(l+ 1 -2 \overline{h}\big) \right)\left(\sum _{l=0}^{2 n} \frac{1}{l+ 1-2 \overline{h}}-\psi \Big(\frac{3}{2} + n\Big)\right) \, ,
\\
&\frac{\partial {\chi _{1}}}{\partial {k}} \Bigg|_{0} = \frac{1}{\sqrt{\pi }} \, \frac{i \, \beta }{2\pi } \, \sum_{n=0}^{\infty} \, \frac{\sec^{2n}\pi (\zeta +\zeta _{c})}{ \Gamma (2n + 2)}\left(\prod _{l=0}^{2 n} \big(l+ 1 -2 \overline{h}\big) \right)\left(\sum _{l=0}^{2 n} \frac{(-)^{l}}{l+ 1-2 \overline{h}}\right) \, ,
\\
&\frac{\partial {\chi _{2}}}{\partial {\omega }} \Bigg|_{0} = \frac{1}{\sqrt{\pi }} \, \frac{i \, \beta }{2\pi } \, \sum_{n=0}^{\infty} \, \frac{\sec^{2n}\pi (\zeta +\zeta _{c})}{ \Gamma (2n + 1)}\left(\frac{1}{2 \overline{h} }\prod _{l=0}^{2 n} \big(l-2 \overline{h}\big) \right)\left(\frac{1}{2 \overline{h} } + \sum _{l=0}^{2 n} \frac{1}{l-2 \overline{h}}-\psi \Big(\frac{1}{2} + n\Big)\right) \, ,
\\
&\frac{\partial {\chi _{2}}}{\partial {k}} \Bigg|_{0} = - \frac{1}{\sqrt{\pi }} \, \frac{i \, \beta }{2\pi } \, \sum_{n=0}^{\infty} \, \frac{\sec^{2n}\pi (\zeta +\zeta _{c})}{ \Gamma (2n + 1)}\left(\frac{1}{2 \overline{h} }\prod _{l=0}^{2 n} \big(l -2 \overline{h}\big) \right)\left(\frac{1}{2 \overline{h} } + \sum _{l=0}^{2 n} \frac{(-)^{l}}{l-2 \overline{h}}\right) \, .
\end{split}
\end{equation}
These expansions can then be recombined in terms of $\xi ^{\mathrm{in}}_{1}$ and $\xi ^{\mathrm{in}}_{2}$ and plugged into the infalling solution. While this procedure is extremely general, there are specific cases when these expansions take a nice form. In particular, for integer twists these resummations can be performed in terms of multiple polylogarithms to \emph{all} orders in $(\omega ,k)$  \cite{Kalmykov:2008ge}. A glimpse of this simplification is already visible from the first order contributions computed above: allowing $2 \overline{h}  = p \in \mathds{Z}^{+}$, the co-efficient of $f(\zeta )^{q}$ for all $2q>p$ is no more a finite sum, but instead collapses to a single term. As a simple case, for $2 \overline{h} =1$ the constituent functions of our infalling solution, $\xi ^{\mathrm{in}}_{1}$ and $\xi ^{\mathrm{in}}_{2}$, can be expanded as,
\begin{equation} 
\begin{split}
&\xi ^{\mathrm{in}}_{1}(\epsilon \omega ,\zeta , \epsilon k) = \frac{\big(1 + \sec \pi (\zeta +\zeta _{c})\big)^{1/2}}{\sqrt{\pi }} \Bigg[1 -  \epsilon \frac{i \beta }{4\pi } \Big( \text{Li}_1(-\sec \pi (\zeta +\zeta _{c})) \, \big(\omega +k\big) + (\gamma +\log 4) \, \omega \Big)+ \mathcal{O}(\epsilon ^{2}) \Bigg],
\\
&\xi ^{\mathrm{in}}_{2}(\epsilon \omega ,\zeta , \epsilon k) = \frac{\big(1 -\sec \pi (\zeta + \zeta _{c})\big)^{1/2}}{\sqrt{\pi }} \Bigg[1 -  \epsilon \frac{i \beta }{4\pi } \Big( \text{Li}_1(\sec \pi (\zeta +\zeta _{c})) \, \big(\omega +k\big) + (\gamma +\log 4) \, \omega \Big)+ \mathcal{O}(\epsilon ^{2}) \Bigg]
\end{split}
\end{equation}
The above expansions lead to the following first order gradient expansion functions,
\begin{equation} 
\begin{split}
&\mathcal{W}_{+}(\zeta ) = \frac{1}{2\pi i} \, \Big( \text{Li}_1(\sec \pi (\zeta +\zeta _{c}))+\text{Li}_1(-\sec \pi (\zeta +\zeta _{c}))+ \log (1+\sqrt{1-\sec ^{2}\pi (\zeta +\zeta _{c})}) + \log 4\Big)
\\
&\mathcal{W}_{-}(\zeta ) = \frac{1}{2 \pi i} \, \Big(\text{Li}_1(\sec \pi (\zeta +\zeta _{c}))-\text{Li}_1(-\sec \pi (\zeta +\zeta _{c}))\Big)
\\
&\mathcal{K}_{+}(\zeta ) =  \frac{1}{2\pi i} \, \Big( \text{Li}_1(\sec \pi (\zeta +\zeta _{c}))+\text{Li}_1(-\sec \pi (\zeta +\zeta _{c}))+ \log 4\Big)
\\
&\mathcal{K}_{-}(\zeta ) = \frac{1}{2 \pi i} \, \Big(\text{Li}_1(\sec \pi (\zeta +\zeta _{c}))-\text{Li}_1(-\sec \pi (\zeta +\zeta _{c})) \Big)
\end{split}
\end{equation}
It is easily checked that these functions satisfy their relevant differential equations. Further, in accordance with \eqref{Horizon values}, the `$-$'  functions have the horizon values,
\begin{equation} 
\begin{split}
\lim_{\zeta \rightarrow \zeta _{h}} \ \frac{1}{\sqrt{f}} \, \mathcal{W}_{-}(\zeta) = \lim_{\zeta \rightarrow \zeta _{h}} \ \frac{1}{\sqrt{f}} \, \mathcal{K}_{-}(\zeta) = \frac{1}{i \pi }.
\end{split}
\end{equation}

\section{Asymptotic expansions of $\mathcal{W}_{\pm}, \mathcal{K}_{\pm}$ }\label{app:Asymp}
In this appendix, we will describe how to derive the asymptotic expansions near the AdS boundary for the $\mathcal{W}_{\pm}, \,  \mathcal{K}_{\pm}$  functions that appear in the massive case.  
Our strategy would be to recast the differential equations obeyed by these functions
in terms of a new variable which then admits a solution in terms of standard functions.
The known asymptotic expansions in terms of incomplete Euler Beta functions can then be used to give a series representation of  these functions near the AdS boundary.

We will begin by defining a new radial co-ordinate,
\begin{equation}
\Upsilon \equiv \fr{1 - \sqrt{f}}{1+\sqrt{f}} \ .
\end{equation}
This coordinate has the limits,
\begin{equation}
\begin{split}
\lim_{r\to \infty} \Upsilon =&\  0 \ , \qquad \lim_{r\to r_h} \Upsilon = 1 \ .
\end{split}
\end{equation}

In terms of this coordinate, the differential equations that occur in the first order in derivative expansion take the form, 
\begin{equation}
\begin{split} \label{WKdefUpsilon}
\fr{d}{d \Upsilon} \mathcal{W}_{+} =&\  \fr{i}{\pi}2^{\fr{2}{d}} \fr{\Upsilon^{\fr{1}{d}-1}}{1-\Upsilon}\fr{\mathcal{F}_{-}^2}{(1+\Upsilon)^{\fr{2}{d}-1}}  \ ,\qquad \quad \;   \fr{d}{d \Upsilon} \mathcal{W}_{-} = \fr{i}{\pi}2^{\fr{2}{d}} \fr{\Upsilon^{\fr{1}{d}-1}}{1-\Upsilon}\fr{\mathcal{F}_{+}\mathcal{F}_{-}}{(1+\Upsilon)^{\fr{2}{d}-1}} \, ,  
 \\ 
\fr{d}{d \Upsilon} \mathcal{K}_{+} =&\  \fr{i}{\pi}2^{\fr{2}{d}} \fr{\Upsilon^{\fr{1}{d}-1}}{{(1+\Upsilon)^{\fr{2}{d}}}}\mathcal{F}_{+}\mathcal{F}_{-}   \, , \qquad \qquad \ \ \fr{d}{d \Upsilon} \mathcal{K}_{-} = \fr{i}{\pi}2^{\fr{2}{d}} \fr{\Upsilon^{\fr{1}{d}-1}}{{(1+\Upsilon)^{\fr{2}{d}}}}\fr{\left(\mathcal{F}_{+}^2 +\mathcal{F}_{-}^2\right)}{2} \, . 
\end{split}
\end{equation}
The solutions are then uniquely fixed by the boundary conditions,
\begin{equation} \label{eq:BdryCondUpsilon}
\begin{split}
\lim_{\Upsilon \to 0-i \epsilon}  ( \mathcal{W}_{+} - \mathcal{W}_{-}) &=0 \, , \qquad \lim_{\Upsilon \to 0-i \epsilon}( \mathcal{K}_{+} - \mathcal{K}_{-}) =0 \, , \\
\lim_{\Upsilon \to 1^{-}} \mathcal{W}_{-} &=0 \, , \qquad \qquad \qquad \lim_{\Upsilon \to 1^{-}}\mathcal{K}_{-} =0 \, .
\end{split}
\end{equation}

The advantage of working in this new co-ordinate is that the above coupled ODEs can be
solved exactly in terms of hypergeometric and Appell functions. We find it convenient to define the following combinations of Appell and hypergeometric functions 
\begin{equation}
\begin{split}
    \mathscr{W}(m,\Upsilon) \equiv &\  4^{\fr{1}{d}-1}\fr{d}{i \pi} \Upsilon^{\fr{1-2m}{d}}   \Bigg\{\fr{{}_{2}F_{1}\left(\fr{2}{d},\fr{1-2m}{d},1+\fr{1-2m}{d},-\Upsilon \right)}{2m-1} 
    \\
    & \qquad \qquad \qquad \qquad \qquad \quad + 2 \Upsilon\fr{AF_{1}\left(1+\fr{1-2m}{d},\fr{2}{d},1,2+\fr{1-2m}{d},-\Upsilon ,\Upsilon\right)}{2m-d-1} \Bigg\} 
    \\
    \mathscr{K}(m,\Upsilon) \equiv &\    \fr{4^{\fr{1}{d}-1}}{i\pi} \fr{d}{2m-1} \;\Upsilon^{\fr{1-2m}{d}} {}_{2}F_{1}\left(\fr{2}{d},\fr{1-2m}{d},1+\fr{1-2m}{d},-\Upsilon \right)
\end{split}
\end{equation}
Using the above, we get the following explicit expressions for the functions that appear in the derivative expansion:
\begin{equation}
    \begin{split}
        \mathcal{W}_{-} =&\   \Bigl[\mathscr{W}(m,\Upsilon) - \mathscr{W}(-m,\Upsilon) \Bigr]^{\Upsilon}_{1^{-}}  \\
        \mathcal{K}_{-}=&\  \Bigl[\mathscr{K}(m,\Upsilon) + \mathscr{K}(-m,\Upsilon)  \Bigr] ^{\Upsilon}_{1^{-}} \\
        \mathcal{W}_{+}=&\  \Bigl[\mathscr{W}(m,\Upsilon) + \mathscr{W}(-m,\Upsilon)-2\mathscr{W}(0,\Upsilon)  \Bigr]^{\Upsilon}_{0} + \lim_{\Upsilon \to 0} \mathcal{W}_{-} \\
        \mathcal{K}_{+}=&\  \Bigl[\mathscr{K}(m,\Upsilon) - \mathscr{K}(-m,\Upsilon) \Bigr]^{\Upsilon}_{0} + \lim_{\Upsilon \to 0} \mathcal{K}_{-}
    \end{split}
\end{equation}
Using the asymptotic expansion,
\begin{equation}
    \begin{split}
        \mathscr{W}(m,\Upsilon) &=\  \fr{4^{\fr{1}{d}-1}}{i\pi}\Bigg(\fr{d}{2m-1} \Upsilon^{\fr{1-2m}{d}} -2^{1-\fr{2}{d}} B\left(1+\fr{1-2m}{d},0,\Upsilon\right)
\\
&\qquad +\Upsilon^{\fr{1-2m}{d}} \sum_{j=1}^{\infty}  \left[ 2^{1-\fr{2}{d}} B\left(j,\fr{2}{d} ,-1\right) - \fr{(-)^{j}}{j} \right]\fr{d}{B\left(j,\fr{2}{d},1 \right)} \fr{\Upsilon^{j}}{1+dj-2m} \Bigg)\ ,
    \end{split}
\end{equation}
we obtain,
\begin{equation}
\begin{split}
\mathcal{W}_{-} &=\  4^{\fr{1}{d}-1} \fr{d}{ i\pi}\Bigg[\fr{\Upsilon^{\fr{1-2m}{d}}}{2m-1} -\fr{2^{1-\fr{2}{d}}}{d} B\left(1+\fr{1-2m}{d},0,\Upsilon\right)
\\
&\qquad \qquad \qquad \qquad \qquad \qquad +\Upsilon^{\fr{1-2m}{d}} \sum_{j=1}^{\infty}  \left[ 2^{1-\fr{2}{d}} B\left(j,\fr{2}{d},-1 \right) - \fr{(-)^{j}}{j} \right]\\
&\ \times \fr{1}{B\left(j,\fr{2}{d},1 \right)} \fr{\Upsilon^{j}}{1+dj-2m}  + \fr{\Upsilon^{\fr{1+2m}{d}}}{2m+1} +\fr{2^{1-\fr{2}{d}}}{d} B\left(1+\fr{1+2m}{d},0,\Upsilon\right) \\
& \qquad \qquad  -\Upsilon^{\fr{1+2m}{d}} \sum_{j=1}^{\infty}  \left[ 2^{1-\fr{2}{d}} B\left(j,\fr{2}{d},-1 \right) - \fr{(-)^{j}}{j} \right]  \fr{1}{B\left(j,\fr{2}{d},1 \right)} \fr{\Upsilon^{j}}{1+dj+2m} \Bigg]^{\Upsilon}_{1^{-}} \, . \\ 
\end{split}
\end{equation}
Since all the variable dependent terms in the above series expansion either vanishe or diverge as \(\Upsilon \to 0 \), we can read of the finite part as the subtracted value at horizon (\(\Upsilon \to 1^{-} \)). 
\begin{equation}
\begin{split}
\mathcal{W}_{c}  =&\ - \fr{d}{ 4\pi i}\Bigg[\fr{ 4^{\fr{1}{d}+1} m}{4m^2-1} +\fr{2}{d} \left\{\psi\left(\fr{d+1-2m}{d}\right)-\psi\left(\fr{d+1+2m}{d}\right)\right\} \\
&\ - \sum_{j=1}^{\infty}  \left[  B\left(j,\fr{2}{d},-1 \right) - 2^{\fr{2}{d}-1}\fr{(-1)^{j}}{j} \right] \fr{1}{B\left(j,\fr{2}{d} ,1\right)} \fr{8m}{4m^2-(1+dj)^2} \Bigg]  \\
\stackrel{d\to 2}{=} &\ - \fr{1}{ 2\pi i}\Bigg[\fr{ 8 m}{4m^2-1} +\psi\left(\fr{3}{2}-m\right)-\psi\left(\fr{3}{2}+m\right) \Bigg] = - \fr{i}{2} \tan m \pi \ .
\end{split}
\end{equation}
where in the last line, we expressed the boundary value of this function for for $d=2$. The infinite series does not contribute since,
\begin{equation}
B\left(j,1,-1 \right) = \fr{(-)^{j}}{j}  \ .
\end{equation}

The function $\mathcal{K}_{-}$ can be written in a manifestly  ingoing way as,  
\begin{equation}
\begin{split}
\mathcal{K}_{-}=&\ \fr{4^{\fr{1}{d}-1}}{i\pi} \Bigl(\mathscr{K}(m,\Upsilon) - \mathscr{K}(-m,\Upsilon^{-1})  \Bigr) \ .
\end{split}
\end{equation}
Expanding this as \( \Upsilon \to 0 \), we get the finite part,
\begin{equation}
\mathcal{K}_{c} = 4^{\fr{1}{d}} \fr{d}{ 4 \pi i} \fr{\Gamma\left(\fr{1+2m}{d} 
	\right) \Gamma\left(\fr{1-2m}{d} 
	\right)}{d \; \Gamma\left(\fr{2}{d} 
	\right)} \ .
\end{equation}
Indeed, this matches the massless limit, 
\begin{equation}
\lim_{m \to 0} \mathcal{K}_{c} = -\fr{i}{2 \sqrt{\pi}} \fr{\Gamma\left(\fr{1}{d} 
	\right)}{\Gamma\left(\fr{d+2}{2d} 
	\right)} \ .
\end{equation}

\newpage
\section{Notation}\label{app:notation}
\small
\begin{center}
	\rowcolors{2}{}{gray!10}
	\begin{tabular} { >{\raggedleft} p {1 cm} | p {5.5 cm} || >{\raggedleft}p {1 cm} | p {5.5 cm} }
		\toprule
		\makebox[1 cm][c]{Symbol} & \makebox[6 cm][c]{Definition} & \makebox[1 cm][c]{Symbol} & \makebox[5.75 cm][c]{Definition} \\
		\midrule
		$d$ &Dimension of the boundary \eqref{AdS BB Metric}. & $\mathcal{Z}$ & Partition function \eqref{SK Partition Function}.  \\
		$\psi $ & Boundary spinor \eqref{AdS CFT fermions}. & $\mathcal{O}$ & Fermionic CFT operator \eqref{AdS Fermion Action}.  \\
		$\Psi $ & Bulk spinor \eqref{AdS Fermion Action}. & $\Delta $ & Conformal dimension \eqref{Conformal dimension}. \\
		$\mathds{S}$ & Action \eqref{AdS Fermion Action}. &$\mathds{S}_{\mathrm{IF}}$ & SK influence functional \eqref{grSK Fermion Action}. \\
		$\mathcal{C}_{\mathrm{SK}}$ & Complex time contour \eqref{SK Partition Function}. & $\mathcal{S}$ & CFT Green's function \eqref{SK Partition Function}.  \\
		\midrule
		$\Gamma ^{(a)}$ & Bulk Gamma  matrices. & $\gamma ^{(k)}$ & Boundary Gamma  matrices. \\
		$\mathbb{\Gamma }$ & $\frac{1}{2} \big(\Gamma ^{(\zeta )} + \Gamma ^{(v)}\big)$ \eqref{BBGamma}. & $\mathcal{P}^{\zeta }_{\pm}$ & Chiral projectors.  \\
		\midrule
		$v$ & Advanced EF time. \eqref{Metric} & $r$ & Radial coordinate \eqref{AdS BB Metric}. \\
		$\bm{x}$ & Transverse spatial co-ordinates \eqref{Metric}. & $r_{h}$ & Radius of the blackhole horizon \eqref{AdS BB Metric}. \\
		$f$ & Emblackening factor \eqref{AdS BB Metric}. & $\beta $ & Inverse temperature of the black brane \eqref{Inverse temperature}. \\
		$\zeta $ & Mock tortoise co-ordinate \eqref{Metric}. & 	$\mathcal{M}_{L,R}$ & Left \& right sheets of the grSK geometry. See below \eqref{Metric}. \\
		$r_{c}, \, \zeta _{c}$ & Finite radial cut-off for $r, \, \zeta $ \eqref{Finite cut-off}. & & \\
		\midrule
		$\omega $ & Frequency conjugate to $v$ \eqref{Fourier transform}. & $\bm{k}$ & Momenta conjugate to $\bm{x}$ \eqref{Fourier transform}. \\
		$E^{(k)}$ & Tetrad 1--forms \eqref{Tetrads}. & $\omega _{ab}$ & Spin connection 1--forms \eqref{Spin connection 1 forms}. \\	
		$m$ & Mass of the fermion \eqref{Dirac Equation}. & &  \\	
		\midrule
		$\mathcal{T}$ & $\mathds{Z}_{2}$ isometry LLT \eqref{PT LLT}. & $\mathfrak{T}$ & $\mathds{Z}_{2}$ isometry spinor LLT \eqref{Spinor PT LLT}. \\
		$\mathcal{J}$ & Jacobian of the GCT \eqref{Tetrads Symmetry}. & $\Psi ^{\mathrm{in}}$ & Infalling solution to \eqref{Dirac Equation FT}. \\
		$\Psi ^{\mathrm{rev}}$ & Frequency--reversed solution \eqref{Psi Out}. & $\Psi ^{\mathrm{out}}$ & Outgoing solution \eqref{Psi Out}. \\
		\midrule
		$S^{\mathrm{in}}$ & Infalling boundary--to--bulk Green's function \eqref{Psi full solution}. & $S^{\mathrm{rev}}$ & Frequency--reversed boundary--to--bulk Green's function \eqref{S Rev}. \\
		$S^{\mathrm{out}}$ & Outgoing boundary--to--bulk Green's function \eqref{S Rev}. & $S_{0}$ & Constant matrix annihilated by $\mathbb{\Gamma }$ \eqref{S0}. \\
		$n^{\mathrm{FD}}_{\omega }$ & Fermi--Dirac factor \eqref{FD factor}. & $c, \, h$& Causal \& Hawking boundary spinors \eqref{Psi full solution}. \\
		$\psi _{L}, \, \psi _{R}$& Left \& right boundary spinors \eqref{psi L R}. & $\psi _{a}, \, \psi _{d}$& Average \& difference boundary spinors \eqref{ad Sources}. \\	
		\midrule
		$H, \, \tilde{H}$ & Functions at $1^{\mathrm{st}}$ \& $2^{\mathrm{nd}}$ order in massless gradient expansion \eqref{H Htilde}. & $\mathcal{F}$& See \eqref{F definition}.\\
		$\mathcal{F}_{\pm} $ & Functions at $0^{\mathrm{th}}$ order in the massive gradient expansion  \eqref{F + - definition}. &	$\mathcal{W}_{\pm}$  & Functions multiplying $\omega $ at $1^{\mathrm{st}}$ order in the massive gradient expansion \eqref{W K definitions}. \\
		$\mathcal{K}_{\pm}$  & Functions multiplying $k_{i}$ at $1^{\mathrm{st}}$ order in the massive gradient expansion \eqref{W K definitions}. & & \\
		\bottomrule
	\end{tabular}
\end{center}

\newpage
\bibliographystyle{JHEP}
\bibliography{FermionicSKholographyV1}

\end{document}